\documentclass[journal]{IEEEtran}
\usepackage{cite}
\ifCLASSINFOpdf
   \usepackage[pdftex]{graphicx}
\else
   \usepackage[dvips]{graphicx}
\fi
\usepackage{amsmath,amssymb,amsfonts}
\usepackage{algorithmic}
\ifCLASSOPTIONcompsoc
 \usepackage[caption=false,font=normalsize,labelfont=sf,textfont=sf]{subfig}
\else
 \usepackage[caption=false,font=footnotesize]{subfig}
\fi
\begin{document}
\title{Towards Designer Modeling through \\Design Style Clustering}
\title{Designer Modeling through Design Style Clustering}

% \author{Michael~Shell,~\IEEEmembership{Member,~IEEE,}
%         John~Doe,~\IEEEmembership{Fellow,~OSA,}
%         and~Jane~Doe,~\IEEEmembership{Life~Fellow,~IEEE}% <-this % stops a space
% \thanks{M. Shell was with the Department
% of Electrical and Computer Engineering, Georgia Institute of Technology, Atlanta,
% GA, 30332 USA e-mail: (see http://www.michaelshell.org/contact.html).}% <-this % stops a space
% \thanks{J. Doe and J. Doe are with Anonymous University.}% <-this % stops a space
% \thanks{Manuscript received April 19, 2005; revised August 26, 2015.}}

\author{Alberto~Alvarez,~\IEEEmembership{Student~Member,~IEEE},
        Jose~Font,
        and~Julian~Togelius,~\IEEEmembership{Member,~IEEE}% <-this % stops a space
\thanks{A. Alvarez, and J. Font are with the Department of Computer Science and Media Technology (DVMT), Malmö University, Malmö, Sweden (e-mail: alberto.alvarez@mau.se; jose.font@mau.se).}% <-this % stops a space
\thanks{J. Togelius is with the Department of Computer Science and Engineering, New York University, New York, NY 11201 USA (e-mail: julian@togelius.com).}% <-this % stops a space
}


\markboth{IEEE Transaction on Games,~Vol.~XX, No.~X, Month~XXXX}%
{Alvarez \MakeLowercase{\textit{et al.}}: Designer Modeling through Design Style Clustering}



\maketitle

\begin{abstract}
We propose modeling designer style in mixed-initiative game content creation tools as archetypical design traces. These design traces are formulated as transitions between design styles; these design styles are in turn found through clustering all intermediate designs along the way to making a complete design. This method is implemented in the Evolutionary Dungeon Designer, a research platform for mixed-initiative systems to create adventure and dungeon crawler games. We present results both in the form of design styles for rooms, which can be analyzed to better understand the kind of rooms designed by users, and in the form of archetypical sequences between these rooms, i.e., Designer Personas. 
% We further discuss how the results here can be used to create style-sensitive suggestions.
% Such suggestions would allow the system to be one step ahead of the designer, offering suggestions for the next cluster, assuming that the designer will follow one of the archetypical design traces.

% We propose modeling designer style in mixed-initiative game content creation tools as archetypical design traces. These design traces are formulated as transitions between design styles; these design styles are in turn found through clustering all intermediate designs along the way to making a complete design. This method is implemented in the Evolutionary Dungeon Designer, a research platform for mixed-initiative systems to create roguelike games. We present results both in the form of design styles for rooms, which can be analyzed to better understand the kind of rooms designed by users, and in the form of archetypical sequences between these rooms. We further discuss how the results here can be used to create style-sensitive suggestions. Such suggestions would allow the system to be one step ahead of the designer, offering suggestions for the next cluster, assuming that the designer will follow one of the archetypical design traces.
\end{abstract}


% \fi
\IEEEpeerreviewmaketitle

\section{Introduction}

% Procedural Content Generation (PCG) is defined as the use of algorithms to generate game content (levels, narrative, visuals, or even game rules) with limited human input \cite{shaker_procedural_2016}. Recent research works have shown PCG's ability to enable new game genres, as well as to create better environments and benchmarks for learning algorithms \cite{risi2019procedural}.

% In parallel, 

% Julian's text

How can we best build a system that lets a human designer collaborate with procedural content generation (PCG) algorithms to create useful and novel game content? Collaboration between AI and humans to co-design and co-create content is a major challenge in AI, and the main focus of Mixed-Initiative Co-Creativity~\cite{yannakakis2014micc,liapis2016mixed}. These systems' objectives are to foster creativity and provide seamless proactive collaboration; ultimately enabling a colleague relation and collaboration as described by Lubart~\cite{LUBART2005-computerPartners}. However, there needs to be an understanding between the human designer and the AI system about what needs to be designed, ideally even a shared goal.

%Collaboration between AI and humans to design and create content is a major challenge in AI, and the main focus of Mixed-Initiative Co-Creative PCG~\cite{yannakakis2014micc}. Multiple approaches have been proposed as alternatives for creating systems that model the interaction between both AI and humans to create game content, and that use different techniques to study such interactions and its implications~\cite{Alvarez2020-ICMAPE,smith_tanagra:_2011,Liapis2014,charity2020baba}. 

% However, for this interaction to be fully fleshed, the human needs to understand the behavior of the AI through interpretable and explainable models and systems, and the AI needs to recognize and interpret the intentions of the humans seamlessly as they create their designs. The former is the focus of~\emph{Interpretable and Explainable AI}, which seek to create or adapt models and systems for a better workflow between humans and AI, where humans can understand the AI's decision process to enable trust relationships and reach deeper interactions~\cite{Zhu2018-XAIDesignersMICC,Doshi-Velez2018,adadi2018peeking}. The latter, which is the focus of this paper, would mean that the AI could adapt its behavior and functionality to the needs, expertise, and workflow of individual designers or specific group of designers. To do so, the AI is required to do an analysis on several design processes such as the designer's preferences, styles, and goals, which holistically is called \emph{Designer Modeling}~\cite{Liapis2013-designerModel,Liapis2014-designerModelImpl}.

Reaching such a shared understanding is a hard task, even when both collaborators share significant cultural and professional backgrounds.
%When one of the collaborators is a computer program, this task is perhaps AI-complete. 
However, we can take steps towards the goal of shared understanding. One idea is to train a supervised learning model on traces of other collaborative creation sessions and try to predict the next step the human would take in the design process. The main problem with this is that people are different, and creators will want to take different design actions in the same state. Another problem is what to do in design states that have not been encountered in the training data. To remedy this, it has been proposed to train multiple models, predicting the next step for different designer ``personas'' (akin to procedural personas in game-playing~\cite{Holmgard2019-proceduralPersonas}). However, for such a procedure to be effective, we need to have sufficient training data. The more different designer personas there are, the more training data is necessary.

One way of overcoming this problem could be to change the level of abstraction at which design actions are modeled and predicted. Instead of predicting individual edits, one could identify different styles or phases of the artifact being created and model how a designer moves from one to another. To put this concretely in the context of designing rooms for a Zelda-like dungeon crawler~\cite{tloz}, one could classify room styles depending on whether they were enemy onslaughts, complex wall mazes, treasure puzzles, and so on. One could then train models to recognize which types of rooms a user creates in which order. By clustering sequences of styles, we could formulate designer personas as archetypical trajectories through style space rather than as sequences of individual edits. For example, in the context of creating a dungeon crawler, some designers might start with the outer walls of the rooms and then populate it with NPCs, whereas another type of designer might first sketch the path they would like the player to take from the entrance to the exit and then add parts of the room outside the main path. These designer models could then be combined with search-based~\cite{Togelius2011} or other procedural generation methods~\cite{khalifa2020-pcgrl} to suggest ways of getting to the next design style from the current one. 

In this paper, we present a method to create and identify designer personas as archetypical paths through style space and provide a prototype implementation of it. For this, we use the Evolutionary Dungeon Designer (EDD), a research platform for exploring mixed-initiative creation of adventure and dungeon crawler content~\cite{alvarez2019empowering,Baldwin2017}. Data from $48$ users designing game levels with the tool following different goals and styles have been used to fit the models. Based on this data, we clustered room styles to identify a dozen distinct types of rooms. To understand the typical progress of designers and validate the clustering, we visualize how typical design sessions traverse the various clusters. We also perform frequent sequence mining on the design sessions to find a small handful of designer personas.

% In this paper, we propose such an alternative approach, designer modeling through clustering the style space, and designer personas as archetypical paths through the clustered style space. We used the Evolutionary Dungeon Designer (EDD), a research platform for investigating mixed-initiative generation of adventure and dungeon crawler game content~\cite{alvarez2019empowering,Baldwin2017}. We analyzed and used data from 190 human-made designs with diverse properties such as multiple goals and style, which in turn, allowed us to identify a set of 12 style clusters as distinct types of rooms. To validate our approach, we used internal validations of multiple clustering approaches and visualized how typical design sessions traverse the clusters. Analyzing this data, and using the steps between clusters of all the 190 designs, we identified a handful of designer personas.

% end of Julian's text

%  designer personas as archetypical paths through style space.

\section{Background}

% Machine Learning (ML) has gained an increased interest from game researchers, achieving remarkable success on training AI agents for very popular games, such as AlphaStar on Starcraft 2 \cite{alphastarblog} and OpenAI Five on Dota 2 \cite{berner2019dota}. 

% Here playing modeling in general
Player modeling, the ability to recognize general socio-emotional and cognitive/behavioral patterns in players~\cite{thawonmas2019artificial}, has been appointed by the game research community as an essential process in many aspects of game development, such as designing of new game features, driving marketing and profitability analyses, or as a means to improve PCG and game content adaptation. Player modeling frequently relies on data-driven and ML approaches to create such models from user data or user-generated gameplay data~\cite{melhart2020feel,Melhart2019-ModellingMotivation,canossa2015towards,Holmgard2019-proceduralPersonas}. 

Modeling users and players can have different goals and use multimodal data. User data can be used to understand and enable behavior for agents in games. For instance, Melhart et al. model a user's \textit{Theory of Mind}, finding that players' perception of an agent's frustration is more a cognitive process than an affective response~\cite{melhart2020feel}. Alvarez and Vozaru explored personality-driven agents, evaluating how observers judged and perceived agents using their personality data from their personality tests when encountering multiple situations~\cite{Alvoz2019-PersonalityDriven}. Gameplay data can show player behavior, as well as help developers understand their player base. Melhart et al. used gameplay data from \textit{Tom Clancy's The Division} to find predictors of player motivation~\cite{Melhart2019-ModellingMotivation}, and Drachen and Canossa used playing behavior data from \textit{Tomb Raider Underworld} and identified four types of players as behavior clusters, which provide relevant information for game testing and mechanic design \cite{Drachen2009-playerModellingTombRaider}.

Furthermore, the combination of Machine Learning (ML) with PCG has led to the rise of Procedural Content Generation via Machine Learning (PCGML), defined as the generation of game content by models that have been trained on existing game content \cite{summerville2018procedural}. PCGML has been used for autonomous content generation, content repair, content critique, mixed-initiative design, or content adaptation. A promising PCGML usage is in the area of content adaptation, where using player and user models are essential to adapt the generated content~\cite{Duque2021-BayesianbasedPlayerModel,togelius2007-AutomaticPersonalisedRaceGames,Yannakakis2011-experiencedrivenPCG}. 

Content adaptation can take place as players play or use the content online or offline, building models from collected data. For instance, Duque et al. adapt and adjust the difficulty of generated content as players play the game using bayesian optimization~\cite{Duque2021-BayesianbasedPlayerModel}. Summerville et al. model players automatically and implicitly by learning from video traces; generating levels that correspond to the latent player models~\cite{Summerville2016-LearningPlayerTailoredPlatformer}. Player models can also be used to enhance and adapt design tools, specifically MI-CC tools. Migkotzidis and Liapis use player models as surrogate models to generate content assisting game designers in the creation of more relevant content for specific players~\cite{Panagiotis2021-susketch}. Similarly, Holmgård et al. use player personas based on player archetypes as content critics to help designers adapt their content to different archetypes~\cite{Holmgard2019-proceduralPersonas}. Their work on player personas is similar to our proposed work, yet instead of personas based on player archetypes, we propose personas based on design style.

\subsection{Designer-centric Perspective}

Mixed-initiative co-creativity (MI-CC)~\cite{yannakakis2014micc}, is the subset of PCG algorithms where human users and AI systems engage in a constant mutual inspiration loop towards the creation of game content \cite{charity2020baba,machado2019pitako,shaker2013ropossum,smith_tanagra:_2011,liapis_generating_2013}. Understanding the designer's behavior and experience, as well as predicting their intentions is key for mixed-initiative creative tools while aiming to offer in real-time user-tailored procedurally generated content. 

While player modeling is key for generating content adapted to players, adapting tools, systems, and AI methods for designing games requires a shift towards a designer-centric perspective, focusing on Designer Modeling. \emph{Designer Modeling}, akin to player modeling, refers to the creation of models of either individual designers or groups of designers informed by how they create various types of content. Liapis et al.~\cite{Liapis2013-designerModel,Liapis2014-designerModelImpl} introduced designer modeling for personalized experiences when using computer-aided design tools, with a focus on the integration of such in automatized and mixed-initiative content creation. The focus is on capturing the designer's style, preferences, goals, intentions, and iterative design process to create representative models of designers.

% to create personalized experiences and adapted tools. Liapis et al. described designer modeling as capturing the style, process, preferences, intentions, and goals of designers as they create content~\cite{Liapis2013-designerModel}. Furthermore, Liapis et al. implemented a prototype of such model in the Sentient Sketchbook~\cite{Liapis2014-designerModelImpl}, where the focus was on using the designer's current design and choice-based evolution to capture process, style, and goals. Alvarez and Font~\cite{Alvarez2020-DesignerPreference}, proposed the use of the behavior characteristics of the generated levels where a neural network was trained on the estimated preference of the designer using a similar approach as the choice-based evolution from Liapis et al.~\cite{Liapis2014-designerModelImpl}.

% adapting tools, systems, and AI methods 

% A paradigm that seeks to combine 

EXplainable AI (XAI) is an emergent research field that holds substantial promise for improving model explainability while maintaining high-performance levels~\cite{adadi2018peeking,Doshi-Velez2018}. Explanations should be aligned with the users' understanding to not hinder the usability of systems, as demonstrated by Nourani et al.~\cite{Nourani2019-meaningfulExplanations}, who discuss the effects of meaningful and meaningless explanations to users of an AI interactive systems.

Zhu et al.~\cite{Zhu2018-XAIDesignersMICC} proposed the field of eXplainable AI for Designers (XAID) as a human-centered perspective on MI-CC tools. This work discusses three principles of mixed-initiative, \emph{explainability}, \emph{initiative}, and \emph{domain overlap}, where the latter focuses on the study of the overlapping creative tasks between game designers and black-box PCG systems in mixed-initiative contexts. This work deems of high relevance the inclusion of data-driven and trained artifacts to facilitate a fluent bi-directional communication of the internal mechanisms of such a complex co-creative process in which \textit{the designer provides the vision, the AI provides capabilities, and they merge that into the creation}. Mapping the designer's internal model to the AI's internal model is suggested as a meaningful way for creating a common ground that establishes a shared language that enables such communication. Our method, and designer modeling in general, aims to develop this designer's internal model to enhance MI-CC tools by adapting the AI's functionality towards designers' needs and aligning to their objectives.

% Recent research proposes visualization techniques based on game design patterns \cite{guzdial2018explainable} and interactive level designer tools \cite{xie2019interactive}, aiming to explain machine learning principles to game designers. 

% Our work  aims at developing these designer's models 

% of designers and align 

Moreover, Guzdial et al.'s~\cite{guzdial-lvldsg-aiide-2018} discuss the insufficiency of current approaches to PCGML for MI-CC, as well as the need for training on specific datasets of co-creative level design. Guzdial et al. work on the mixed-initiative Morai Maker~\cite{guzdial2019friend} shows the relevance of exploring the ways designers and AI interact towards co-creation, identifying four human-AI relationships (friend, collaborator, student, and manager), as well as the different ways they impact on the designer-user experience. Our study advocates for the importance of designer modeling through ML as the generation of surrogate models of designer styles by training on existing designer-generated data, aiming for an improvement in quality and diversity in computational creativity and, in particular, MI-CC tools.

\subsection{The Designer Preference Model in EDD}

EDD is an MI-CC tool where designers can create dungeons and rooms; meanwhile, a PCG system analyzes their design and proposes suggestions to the designer~\cite{Alvarez2018, Baldwin2017}. EDD uses \emph{Interactive Constrained MAP-Elites} (IC MAP-Elites)~\cite{alvarez2019empowering}, an evolutionary algorithm that combines Constrained MAP-Elites~\cite{Khalifa2018} with interactive and continuous evolution. 

The work presented in \cite{Alvarez2020-DesignerPreference} introduced the Designer Preference Model, a data-driven solution that learns from user-generated data based on their choices while using EDD. Both systems constantly interact and depend on each other, so that the Designer Preference Model learns from the selected suggestions, and IC MAP-Elites uses the Designer Preference Model as a designer's surrogate model to complement the fitness evaluation of new individuals. The results showed the need for stability and robustness in the data-driven model, to counterbalance the highly dynamic designer's creative process.  

% This approach's main goal is modeling the user's design style to better assess the tool's generated content, increasing the user's agency over the generated content without stalling the MI-CC loop \cite{ComptonPhD} or decreasing user fatigue with periodical suggestion handpicking \cite{liapis2016mixed,Takagi2001-InteractiveEvo}. The results showed the need for stability and robustness in the data-driven model, to counterbalance the highly dynamic designer's creative process. 

\begin{figure*}[ht!]
\centerline{\includegraphics[width=\textwidth]{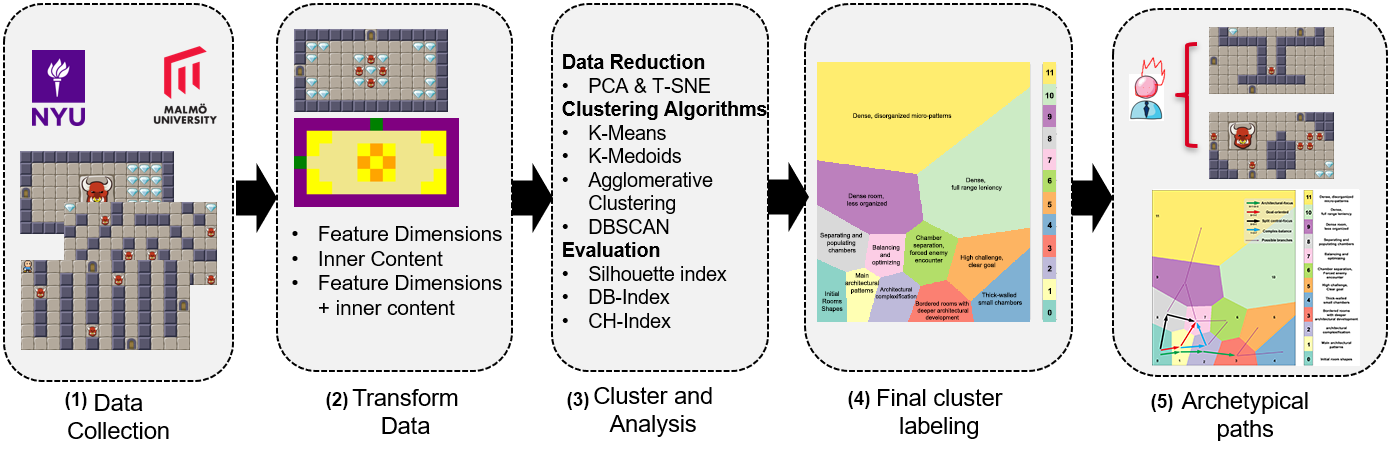}}
\caption{The stages of the design style clustering development: (1) Data was first collected through two user studies. (2) Then, using the design sequences, the data was processed into five different datasets, one using the room images, a second using the tiles information, and three using tabular information. (3) A data reduction technique was applied to different datasets, and then they were clustered and internally evaluated. (4) The clusters were formed, picked from the best performing methods, and labeled based on the data points within each cluster. The clusters were evaluated by visualizing how a typical design session traverse the various clusters, and K-Means (K=12) was chosen as the final approach. (5) Finally, using this final approach all the sequences were clustered and archetypical paths were identified.%(5) The final approach,  K-Means (K=12) was evaluated by visualizing how a typical design session traverse the various clusters. Finally, the sequences were clustered by the final approach and archetypical paths were identified.
} \label{fig:approach-steps}
\end{figure*}
\section{Concepts and Definitions}

%This paper presents an approach and fundamental steps towards the implementation of designer personas: an analysis of designer style clustering to isolate archetypical paths that can be later be used to build ML surrogate models of archetypal designers. Such models would adapt to the dynamic designer during the mixed-initiative creative process by being placed in the solution space, allowing the designer to traverse such space of models as she drifts through the many dimensions of her creative process.

% design archetypes 

Our work draws from ideas, concepts, and definitions introduced by Liapis et al., such as the core designer model loop when using CAD tools, what can be modeled: preferences, style, goals, processes, and their definition, and particularly, the use of designer modeling as an individual or collective model~\cite{Liapis2013-designerModel}. We support our approach on the idea of style as a particular type of designer's preference, and that a collective model can be used to form a stable and static design space, which after being interacted with by designers, can be adapted towards them.

% and the idea of style as a particular type of designer's preference.

%. Moreover, Liapis et al. discuss the modeling of style as a type of preference, where each individual designer has peculiarities and characteristics that makes their style recognizable. While we agree with this vision, we 

%Our work draws from many of the ideas and concepts introduced by Liapis et al.~\cite{Liapis2013-designerModel}, in relation to style, goals, preferences and design processes of designers. Nevertheless, given the interdisciplinary scope of this system, and the multiple concepts discussed throughout the paper, it is essential to have operational definitions on the different terms used.

%Thus, in this paper, the shared goal is set and defined by the designer with her design, and as she develops, adapts, and changes, the system seeks to adjust its goals to support the designer's work. Furthermore, the aim of this paper is to propose a system that is able to identify the designer's current goal and style to adapt further the system's goals to provide a personalized experience.

\subsection{Design Style} \label{sec:designStyle}

%Every designer has a different style when creating content, especially levels, where one might 

% One idea is to train a supervised learning model on traces of other collaborative creation session and try to predict the next step the human would take in the design process. The main problem with this is that people are different, and different creators will want to take different design actions in the same state;

% One way of overcoming this problem could be to change the level of abstraction at which design actions are modeled and predicted. Instead of predicting individual edits, one could identify different styles or phases of the artifact being created, and model how a designer moves from one to another. To put this concretely in the context of designing rooms for a Zelda-like dungeon crawler~\cite{tloz}, one could classify room styles depending on whether they were enemy onslaughts, complex wall mazes, treasure puzzles, and so on. One could then train models to recognize which types of rooms a user creates in which order. By clustering sequences of styles or phases we could formulate designer personas as archetypical trajectories through style space, rather than as sequences of individual edits. For example, in the context of creating a dungeon crawler, some designers might start with the outer walls of the rooms and then populate it with NPCs, whereas another type of designer might first sketch the path they would like the player to take from the entrance to the exit and then add parts of the room outside the main path.

There exist many different styles when creating content, especially levels, that designers can create and adapt to accomplish their goals and the experiences they want for players. On a general level, \emph{Design Style} encompasses the creative process from conceptualization, prototyping, reflection, adaptation, especially when following different processes or constraints during collaboration. Taking a more concrete and operational level, \emph{Design Style} can be analyzed as overarching goals that different designers have when creating a dungeon. For instance, dungeons in games such as Zelda\cite{tloz} or The Binding of Isaac\cite{mcmillen_binding_2011}, represent a particular playing style planned by the designer. In the former, low tempo, exploring the dungeon, and secret rooms define the style of the dungeons, whereas in the latter, high tempo, optimizing time and resources, small rooms, and in general high-challenge define the dungeons. 

While interesting and relevant to understanding the designers' holistic design process and the expected player experience, \emph{Design Style} can also be discussed on an individual room basis. Rooms have their own set of characteristics and styles that can be identified and modeled to understand their design process. Some would prefer to create the room's architecture first to then create the goals within, whereas others would like to place strategic objectives around and then create the architecture around it or alternating between both. Even with such a division, how to reach those design styles is not straightforward and does not require the same strategy, which also shows the preference and style of individual designers. For instance, if the goal is to create a challenge to reach a door, the designer could create a room with a substantial number of enemies, create a concentrated high-challenge in the center of the room, or divide the room into smaller choke areas. Therefore, in this paper, we take a simplified view of \emph{Design Style} and treat it as the style designers follow to create a room, informed by the individual steps each has taken connected to their preferences and goals.

\section{Room Style Clustering}
% \begin{enumerate}
%     \item Information on the user studies and how we develop the clusters.
%     \item collected data through 2 user studies
%     \item transformed the data into 5 datasets
%     \item data reduction through PCA and T-SNE. Cluster with K-means, K-Medoids, agglomerative clustering and DBSCAN. Evalulated through internal indices (silhouette index, DB index, and CH-index) 
% \end{enumerate}

\begin{figure*}[t]
\centerline{\includegraphics[width=15cm]{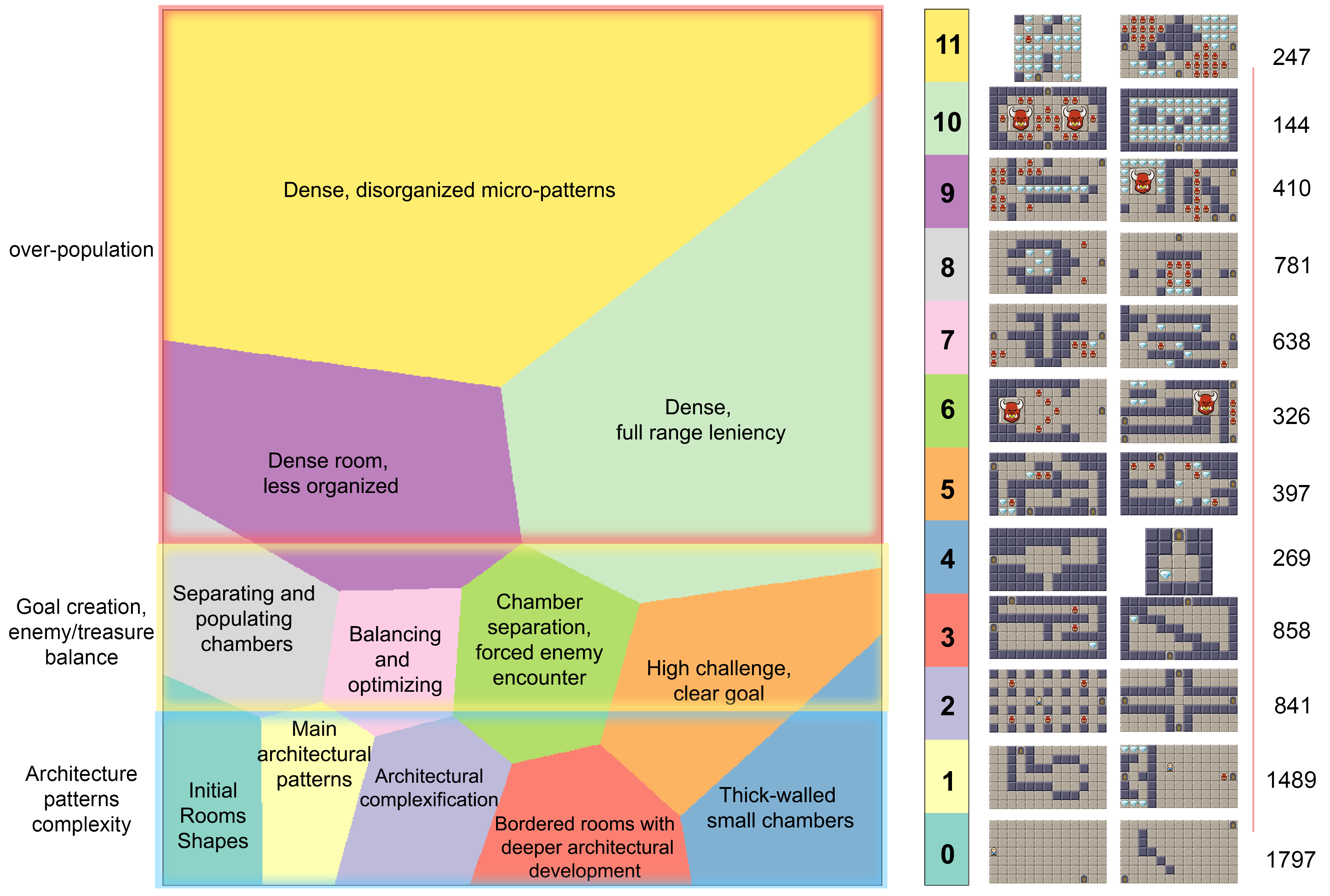}}
\caption{Best resulting cluster set. K-Means (K=12), using the \textbf{Tiles} Dataset. While it scores slightly less in the internal indices that other setups, a qualitative analysis successfully gives us more granularity by subdividing the main bottom clusters, to label and cluster the design process of designers. Sample rooms belonging to each cluster are displayed on the right, next to the total number of rooms in the cluster.} \label{fig:all-clusters}
\end{figure*}

% \begin{figure*}[b]
% \centerline{\includegraphics[width=13cm]{figures/representative cluster-steps.png}}
% \caption{Examples of a step by step edition sequence of a design session and it's clustering. To the left, we present the actual sequence and steps of one of the rooms in the dataset and to the right is the actual trajectory of the design in the cluster space. Numbered and in black, it is shown how each step of the design process is clustered by our approach} \label{fig:paths-designers}
% \end{figure*}

% \begin{figure}[h]
% \centerline{\includegraphics[width=8cm]{figures/representative-cluster-steps-alter.png}}
% \caption{Example of a step by step edition sequence of a design session and it's clustering. At the top, we present the actual sequence and steps of one of the rooms in the dataset, in a $4\times7$ grid, starting at the top left with the first edition. At the bottom, it is the actual trajectory of the design in the cluster space. Numbered and in black, it is shown how each step of the design process is clustered by our approach} \label{fig:paths-designers}
% \end{figure}

\begin{figure}
\centerline{\includegraphics[width=0.4\textwidth]{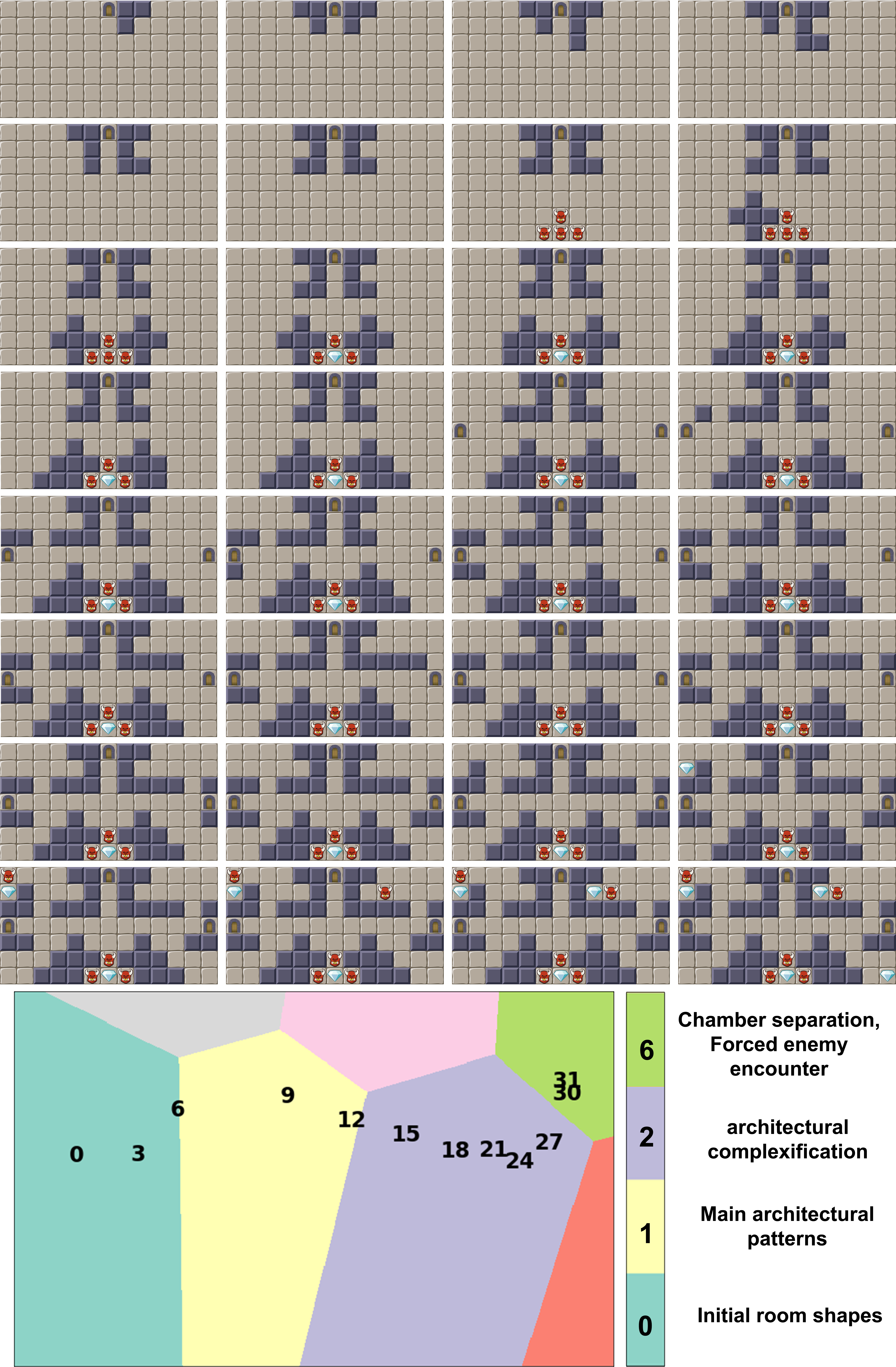}}
\caption{
% Example of a step by step edition sequence of a design session and it's clustering. At the top, we present the actual sequence and steps of one of the rooms in the dataset, in a $4\times7$ grid, starting at the top left with the first edition. At the bottom, it is the actual trajectory of the design in the cluster space. Numbered and in black (every 3 steps), it is shown how each step of the design process is clustered by our approach
Example of a room's design step sequence and its clustering. At the top, we present the design steps of one of the rooms in the dataset starting at the top left with the first design step. At the bottom, it is the actual trajectory of the design in the cluster space. Numbered and in black (every 3 steps), it is shown how each step of the design process is clustered by our approach} \label{fig:paths-designers}
\end{figure}

\begin{table}
\begin{center}
{\caption{Best performing setups based on their internal validation and visualization of clustered data points.}\label{table:setups}}
\resizebox{0.48\textwidth}{!}{
\begin{tabular}{cccccccc}
\hline
\rule{0pt}{12pt}
Algorithm&Data&K&$\Diamond$&$\Box$&$\bigtriangleup$&$ARI$&$ARI(R)$
\\ 
\hline
\\[-6pt]
K-Means & Tiles-PCA & 9 & 0.43 & 0.73 & 9437.23 & 0.941 & 0.581\\ 
$\star$K-Means & Tiles-PCA & 12 & 0.41 & 0.77 & 9436.57 & 0.935 & 0.775\\
K-Means & Dimensions-PCA & 12 & 0.43 & 0.73 & 7738 & 0.84 & 0.876\\
K-Means & Combined-PCA & 10 & 0.41 & 0.73  & 8455.34 & 0.931 & 0.866 \\ 
Agglomerative compl. & Tiles-PCA & 6 & 0.501 & 0.64 & 4221.34 & 0.719 & 0.419\\  
\hline
\\[-6pt]
\multicolumn{8}{l}{$\Diamond$ Silhouette Score\ \
$\Box$ Davies Bouldin Index\ \
$\bigtriangleup$ Calinski-Harabasz Index}
\end{tabular}
}\end{center}
\end{table}

This paper presents an approach and method for the implementation of designer personas: an analysis of designer style clustering to isolate archetypical paths that can later be used to build ML surrogate models of archetypal designers. Such models would allow for an stable and static design and solution space to be traversed by designers, adapting it to them as they explore their creative process.

% Such models would adapt to the dynamic designer during the mixed-initiative creative process by being placed in the solution space, allowing the designer to traverse the space of models as she drifts through the many dimensions of her creative process.

The proposed system builds on top of EDD's Designer Preference Model and preliminary results \cite{Alvarez2020-DesignerPreference}, expanding it to classify the designers' designs based on clusters developed using previously hand-made design sequences by expert and non-expert designers. Figure \ref{fig:approach-steps} illustrates our approach in five sequential stages, from data collection to experimentation and results. The first four stages are explained in the following subsections, whereas Section~\ref{section:results} shows the experimental results.

\subsection{Data Collection}

% We conducted two user studies where participants were tasked with freely designing a dungeon in EDD and the rooms that compose it with no further restrictions,
We conducted two user studies where participants were tasked with designing and creating a dungeon with interconnected rooms without restrictions, using all the available tiles i.e. floor, wall, treasure, enemy, and boss tiles. All participants were introduced to the tool before the design exercise. User-generated data was gathered during the complete design session, creating a new data entry every time the designer edited the dungeon. In total, we had $40$ participants, $25$ of these (i.e. NYU participants) were industry or academic researchers within the Games and AI field, and the other $15$ (i.e. MAU participants) were game design students. This resulted in a diverse dataset composed of $180$ unique rooms like the ones depicted in Figure~\ref{fig:approach-steps}, that was pre-processed and clustered in the subsequent stages. 

\subsection{Dataset pre-processing}

From the $180$ unique rooms, we extracted and used each room's design step sequence, from their initial design to the more elaborated end-design, to compose a richer dataset that could capture the design process of a designer rather than focusing on the end-point. This resulted in a dataset with $8196$ data points. 
% Through this, we ended up using $8196$ data points in our dataset.
%just the end-point. We ended up using $8196$ rooms
Moreover, five different copies of the dataset were created to analyze and compare the performance of the clustering stage using the following image pre-processing methods:
\begin{enumerate}
    \item \textbf{Room:} No pre-processing. Room images are fed into the next stage as they were created by the designer, with a resolution of $1300\times 700\times3$, corresponding to width, height, and RGB ($3$ color channels).
    
    \item \textbf{Tiles:} Each room tile type is mapped to a single-color pixel and the rooms are simplified to a pixel-tile based representation, as shown in the second stage of Figure \ref{fig:approach-steps}. The dimensions are downscaled to $13\times 7\times3$.
    %Each room tile is simplified to a single-color pixel, as shown in the second stage of Figure \ref{fig:approach-steps}, downscaled to $13\times 7\times3$.
    
    \item \textbf{Dimensions:} Each room is described by its five IC MAP-Elites feature dimension values, excluding the similarity scores: \textsc{Linearity}, \textsc{Leniency}, \textsc{\#MesoPatterns}, \textsc{\#SpatialPatterns}, and \textsc{Symmetry}. A complete description of these features can be found in~\cite{Alvarez2020-ICMAPE}.
    
    \item \textbf{Inner Content:} Each room is described by $12$ values, related to the count, sparsity, and density of the enemy, treasure, floor, and wall tiles contained in it.
    
    \item \textbf{Combined:} A combination of the \textbf{Dimensions} and \textbf{Inner Content} methods.
\end{enumerate}

\subsection{Clustering and Analysis}

% To run all setups, data reduction algorithms, clustering algorithms, and do the internal evaluation of the clusters, we used scikit-learn machine learning toolset~\cite{scikit-learn}. 

We used the Scikit-learn machine learning toolset~\cite{scikit-learn} to run all setups, data reduction and clustering algorithms, and to internally evaluate the clusters. To obtain the best set of clusters, we ran different setups with the above datasets. The data was reduced to two meaningful dimensions with two different data reduction algorithms, Principal Component Analysis (PCA) and T-Distributed Stochastic Neighbor Embedding (T-SNE). For both dimension reduction algorithms, we fit the algorithms with each individual dataset, setting to two principal components and in the case of T-SNE using PCA as an initializing algorithm, and transforming the data into a \emph{pca\_dataset} and \emph{tsne\_dataset}. Each two-dimensional point in the new datasets represents a step in the sequences described above. Preliminary tests without dimensionality reduction using the same distance metrics were conducted for all datasets except for \textbf{Room} and \textbf{Tiles} since the amount of dimensions made it computationally infeasible. These did not result in good space partition, achieved worse internal indices, and produced disjointed clusters. For detailed information, see the supplementary material.  % and reducing to $95\%$ of variance within datasets were conducted and resulted in worse internal indices, and disjointed space partition. %Likewise, for the T-SNE, we fit the algorithm with each individual dataset, setting the parameters to two principal components and using PCA as initializing algorithm, and then transformed the data into a new dataset \textit{tsne_dataset} per dataset.

All the resulting datasets were then clustered using \textsc{K-Means, K-Medoids, Agglomerative clustering}, and \textsc{DBSCAN}. K-Means was initialized using the standard k-means++ implementation in scikit-learn, which initializes all centroids distant from each other. K-Medoids was initialized similarly, using the standard k-medoids++, and tested using the \emph{cosine}, \emph{euclidean}, and \emph{manhattan} distances. Agglomerative clustering is a hierarchical clustering approach using a bottom-up approach implemented in scikit-learn using four different linkage criteria for comparing data points: \emph{Ward}, \emph{Complete}, \emph{Average}, and \emph{Single}. Finally, DBSCAN clusters points based on density separated by low-density areas; thus, DBSCAN automatically finds $k$ based on two parameters, $\epsilon$ describing the maximum distance between points and \emph{min\_samples} describing the minimum amount of samples within a group to be considered a cluster. K-Means, K-Medoids, and Agglomerative clustering were tested using multiple $K$ values ranging from 3 to 13, and DBSCAN was tested with several $\epsilon$ values ranging from 0.3 to 1.0, and \emph{min\_samples} ranging from 2 to 9.

%, testing with $K$ values ranging from 3 to 13 for the first three ones, and several $\epsilon$ values for DBSCAN.

%testing different minimum distance between data points ($\epsilon$) and the minimum amount of data points within a cluster to be considered a dense region for DBSCAN.

Since we lack a labeled dataset (i.e. ground truth) for cluster validation, we evaluated the results from all setups using the internal indices below. However, good values in internal indices are indicative and do not imply the best information retrieval, which is why we also manually inspect the clusters and the rooms grouped together.

%Since in our approach lacks a labeled dataset (i.e. ground truth) for cluster validation, 

\begin{itemize}
\item \textbf{Silhouette Score ($\Diamond$):} The Silhouette Score shows how similar a data point is to the cluster it is associated with, through calculating the difference between the $\overline{distance}$ from the point to the points in the nearest cluster and the $\overline{distance}$ to the points in the actual cluster. The value is bounded from -1 to +1, with values closer to +1 indicating a good separation of the clusters, and closer to -1 meaning that some points might belong to another cluster.
\item \textbf{Davies-Bouldin Index ($\Box$):} The DB-index is the ratio between the within-cluster distances and between-clusters distances. With this, we can have an insight into the average similarity of clusters with their closest cluster. The value is bounded from 0 to +1, where values closer to 0 relate to clusters that are farther apart from each other and less dispersed, thus, this index is more crucial when we have more dense representations.
\item \textbf{Calinski-Harabasz Index ($\bigtriangleup$):} The CH-index is another index related to the density of the clusters and how well separated they are. The score is the ratio between the within-cluster dispersion (compactness) and the between-cluster dispersion (separation). The CH-index is positively unbounded, and the higher the score the better.
\item \textbf{Adjusted Rand Index ($ARI$):} $ARI$ computes the similarity between two set of clusters. We tested our setups by using 70\% of the data as training set to fit the cluster algorithm and the other 30\% to analyze how the fitted clusters clustered unseen data. Then, using our setups fitted with the full dataset, we compared these 30\% test set using $ARI$, giving us an estimated performance value with unseen data. Additionally, we computed $ARI(R)$ by splitting the data based on the room sequences rather than on the individual design steps to remove any relation between training and test set.

% (i.e., the possibility for data in the test set to be part of the same room's design step sequence than in the training set).

% the which resulted in instead of computing $ARI$ with random we tested $ARI$ separating the data 
% \item \textbf{Adjusted Rand Index ($ARI$):} $ARI$ computes the similarity between two set of clusters considering what data points were assigned the same cluster or not. In our case, we tested our setups by using 70\% of the data as training set to fit the cluster algorithm and 30\% of the data to analyze how the fitted clusters clustered unseen data. Then, using our setups fitted with the full dataset, we compared these 30\% test set using $ARI$, giving us an estimated value of how the algorithm with the same hyperparameters and similar data would perform. 
\end{itemize}

%  The final column, $ARI$ (Adjusted Rand Index) computes the similarity between two clustering considering what data points were assigned the same cluster or not.
\subsection{Cluster Labelling}

Table~\ref{table:setups} shows the best performing setups according to their internal indices scores and manual qualitative analysis of the clusters, rooms clustered together, and sequence analysis (further explained in section~\ref{section:results}). The algorithm selection process followed a systematic analysis of each algorithm with each dataset from the lowest to the highest hyper-parameter value for each algorithm. We focused on the setups that both scored the best in the different indices (as a combination) and using the \textit{elbow method}, which is a common heuristic in clustering to select the best $k$. We filtered our results by $k\ge6$ as it  gave more meaningful partitions and better groups overall. Some setups with lower $k$ values gave good overall internal indices scores but logically, tended to make big clusters and supersets. For instance, K-Means (k=3) using \textbf{Tiles} dataset scores $\Diamond = 0.55$, $\Box = 0.66$, and $\bigtriangleup = 9049.65$. However, the three clusters acted as supersets, dividing the space into an upper "over-population" cluster, and into two smaller bottom partitions, one to the left accounting for the architecture of the rooms and one to the right focused on challenge and leniency. By using $k\ge6$, we guarantee that at least the big clusters would be divided into meaningful subsets such as the ones shown in figure~\ref{fig:all-clusters}.

When using the \textbf{Dimensions} and \textbf{Combined} datasets, the clusters perform well in certain indices than when using the \textbf{Tiles} dataset with certain algorithms. However, when analyzing the resulting setups, they missed a clear relationship between the clustered rooms, which was exacerbated when analyzing sequences on these setups, where they missed continuity between clusters and either jump around clusters or had close to no movement between them. When using these datasets and analyzing the sequences, we notice that there are sequences where the cluster set is robust to seemingly meaningless changes, whereas jumps occurred when something important changed (or was added), and in other sequences, the clusters are very sensitive to changes. We believe this is because similar rooms (one or a few steps in difference) could have a big impact on several dimensions within these datasets, such as a room becoming more symmetric and fully linear by simply adding a few walls. This, combined with the lack of natural continuity in many of the analyzed examples, negatively affected their evaluation.

Conversely, given that we are creating tile-based rooms and dungeons, the \textbf{Tiles} dataset had more representative features, which when used, generally performed better in the evaluated internal indices, and the clusters meaningfully separated the data. Similar robustness and beneficial behavior as when using the \textbf{Dimensions} and \textbf{Combined} datasets was also encountered while showing a clear continuity between designs that supports its usability (further discussed in section~\ref{section:results}). Figure \ref{fig:all-clusters} shows the best-resulting cluster set found among all the experiments run. K-Means (K=12) with the \textbf{Tiles} dataset was chosen as the best cluster set. While K-Means (K=9) performs better in all indices, K=12 gave us more granularity within the bottom clusters. In our case, this was more important, especially since the difference was minimal, as the bottom area of the design space is denser than others containing much of the early design steps and their refinement.

In the figure, we have plotted on top of the clusters the labels describing in general, the content that is within them. Labels were defined based on the rooms that were clustered together and surrounding clusters' content. The following is a description of the clusters and rooms clustered together:

\textbf{0. Empty-Initial rooms:} %This cluster contains $1797$ data points, and 
This cluster relates mostly to the initial designs made by the designers. These designs are from completely empty rooms to initial work-in-progress structures.

\textbf{1. Main architectural shapes:} Similar to other clusters within the same layer, this cluster relates to the development and definition of main architectural patterns that are somewhat symmetric.

\textbf{2. Architectural complexification:} %This cluster contains $841$ data points, and 
This cluster relates mostly to the complexification of wall structures by having dense wall chunks, representative architectural patterns, or symmetrical patterns.

\textbf{3. Bordered rooms with deeper architectural development:} This cluster relates mostly to rooms with an added wall border by the designer, and where the focus is to shape chambers and develop more visual structures.

\textbf{4. Thick-walled small chambers:} This cluster to the far right region in fig.~\ref{fig:all-clusters}, relates to more highly-linear, confined and maze-like rooms.% with more structure on what is possible. %The cluster contains $269$ data points.

\textbf{5. High challenge, clear goal:} This cluster relates to well-shaped rooms with clear wall structures and goals, towards more challenge. 

\textbf{6. Chamber separation with forced enemy encounter:} This cluster relates to rooms that are in the process of a clear segmentation into corridors and chambers, and that enforce to some extent enemy encounters for the player. 

\textbf{7. Balancing and optimizing:} This cluster contains a mix between corridors and chambers within rooms with a focus on balancing rooms and optimizing their design towards certain goals. %The cluster contains $638$ data points.

\textbf{8. Separating and populating chambers:} This cluster relates to the process of separating rooms into distinct chambers, focusing on the center of the room, and starting to populate rooms with enemies and treasures. %The cluster contains $781$ data points.

\textbf{9. Dense, less organized:} This cluster contains rooms that still have a certain objective but are moving towards more disorganized distributions of micro-patterns in relation to their density. %This cluster contains $410$ data points.

\textbf{10. Dense, full range leniency:} Focusing on density as the other two clusters within the same layer, this cluster relates to rooms that are in the full range of leniency from very rewarding treasure rooms to very challenging boss rooms. %This cluster contains $144$ data points.

\textbf{11. Dense, disorganized micro-patterns:} This cluster contains the extreme rooms that contain a high density of tiles, other than floor-tiles, without a clear structure or objective for the player.

Moreover, besides the local relation between clusters, the clusters are implicitly divided in three layers on the Y-axis. From bottom to top, (a) architectural patterns complexity, relating to clusters composed of rooms with clearer or complex shapes created with walls, from empty rooms to chambers. (b) Goal creation, enemy/treasure balance, with clusters containing the strategic addition of enemies and treasures to establish objectives in the room for the player. In terms of EDD, these rooms are composed of more meso patterns. And (c), over-population, which relates to clusters filled with less organized and dense rooms where the addition of enemy or treasure does not necessarily need to follow any clear objective. Identifying the designer in a layer, and the path they have taken to get there could show meaningful information in the design process. For instance, the intentions of the designer or in what phase of the design process they are at the moment; i.e. trying the tool or observing how the tool reacts or scraping their current goal towards a new goal within the room. 

% \begin{enumerate}
% \setcounter{enumi}{-1}
% \item[0] \textbf{Empty/Initial rooms:}
% \item \textbf{Complex wall mazes:} 
% \item \textbf{Dense, less structure} 
% \item \textbf{Structural complexification:} 
% \item \textbf{Dense, full range leniency} 
% \item \textbf{Separating and populating chambers:} 
% \item \textbf{Balancing and optimizing rooms:} 
% \item \textbf{Bordered rooms with deeper structural development:}
% \item \textbf{Development of main structural shapes:} 
% \item \textbf{Dense, unstructured:} 
% \item \textbf{Challenging rooms with clear goal:} 
% \item \textbf{Chamber separation with forced enemy encounter:} 
% \end{enumerate}

\begin{figure}[t!]
\centerline{\includegraphics[width=9cm]{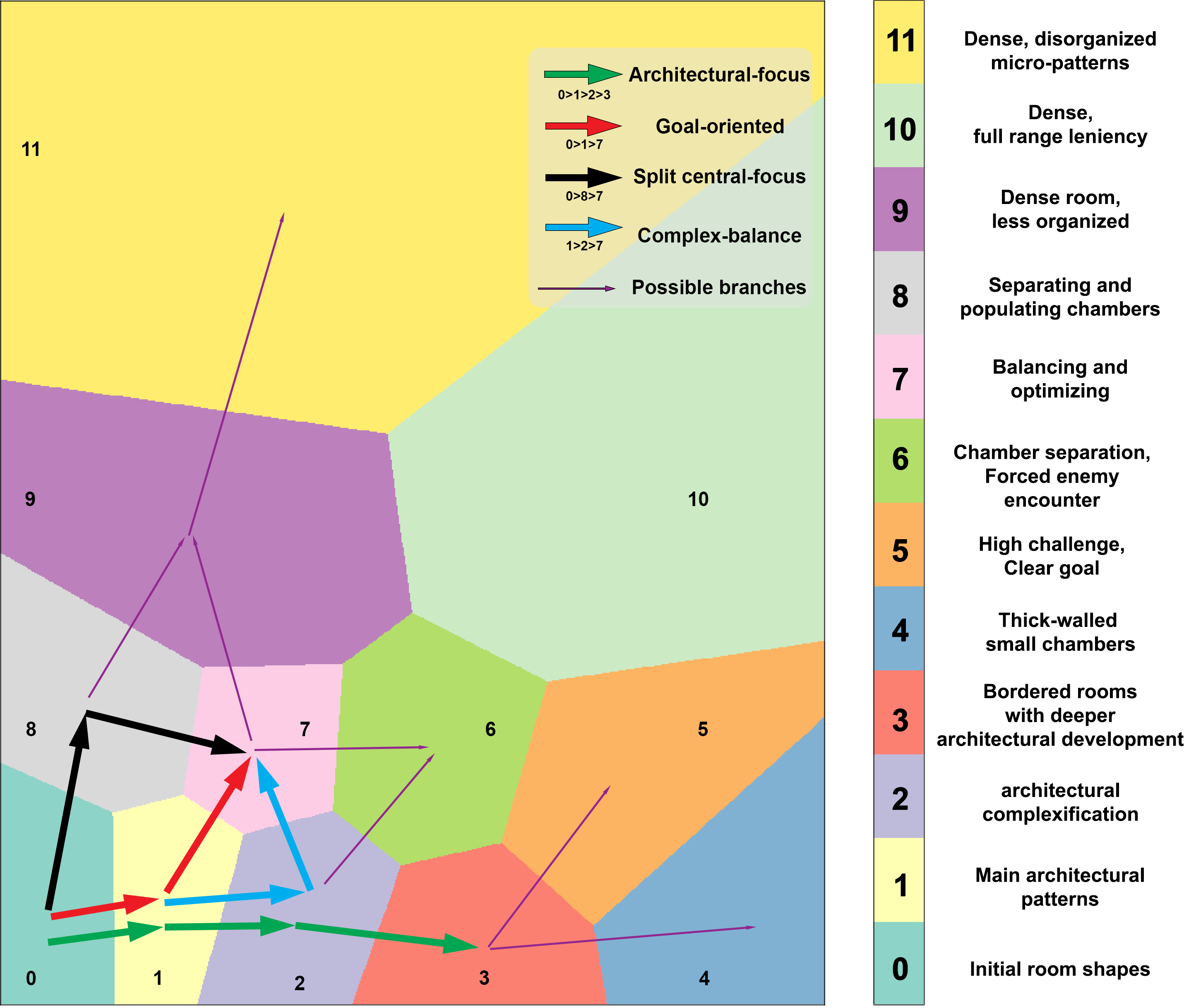}}
\caption{Final and common designer trajectories. The archetypical paths are represented with thick arrows, and were calculated using the frequencies of subsequences from $180$ diverse rooms. Each color represent a unique trajectory; with green the \textsc{Architectural-focus}, with red the \textsc{Goal-oriented}, with black the \textsc{Split central-focus}, and with blue the \textsc{Complex-balance}. Finally, thinner purple arrows extending from clusters traversed by the archetypical paths show the multiple possible branches that an archetypical path can deviate or extend to.} \label{fig:finalPaths}
\end{figure}

\section{Designer Personas} \label{section:results}

\begin{figure}[t!]
    \centering
     \subfloat[\textsc{Architectural-focus}\label{fig:af}]{%
       \includegraphics[width=0.38\textwidth]{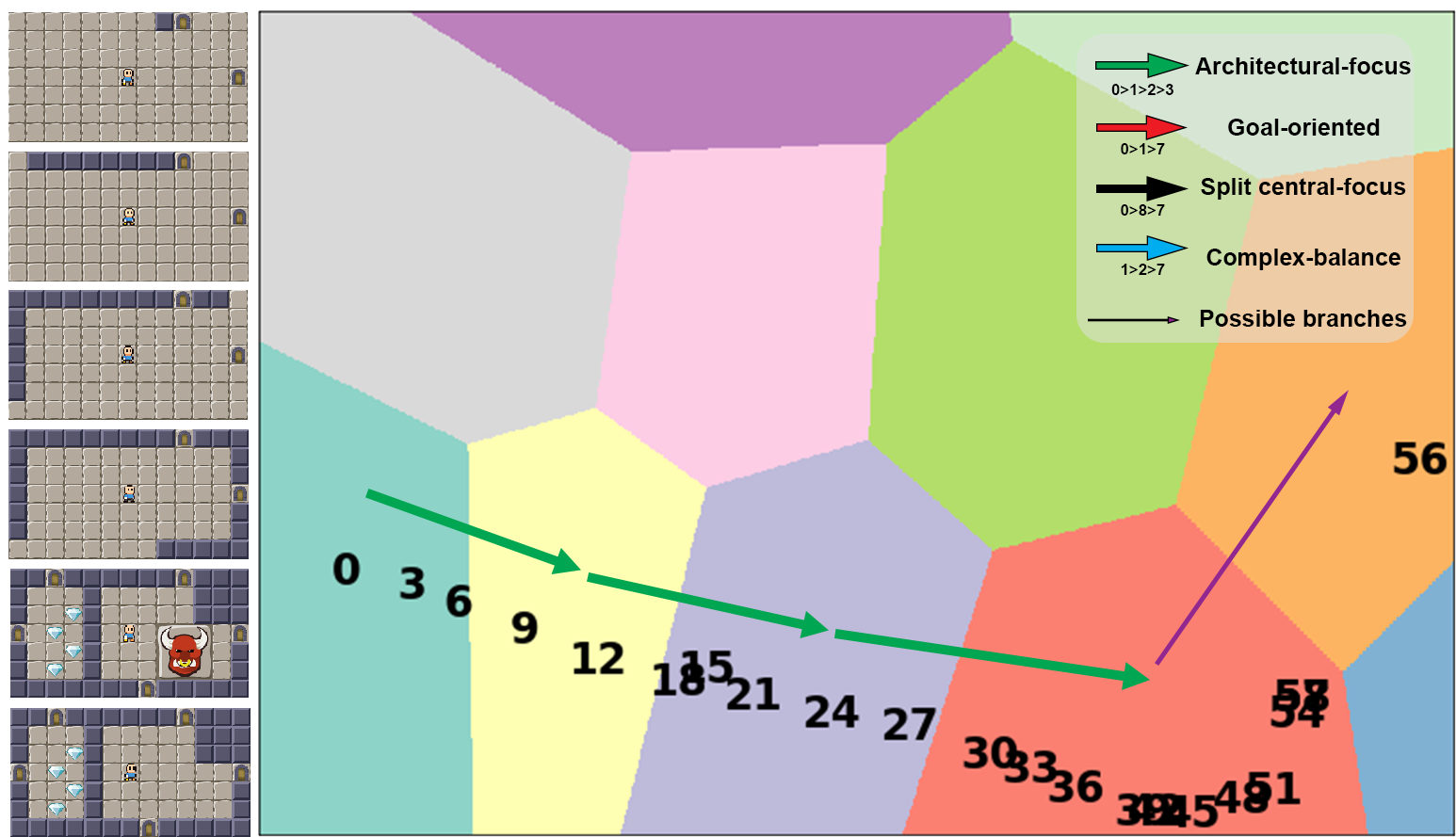}
     }
     \hfill
     \subfloat[\textsc{Goal-oriented}\label{fig:go}]{%
       \includegraphics[width=0.38\textwidth]{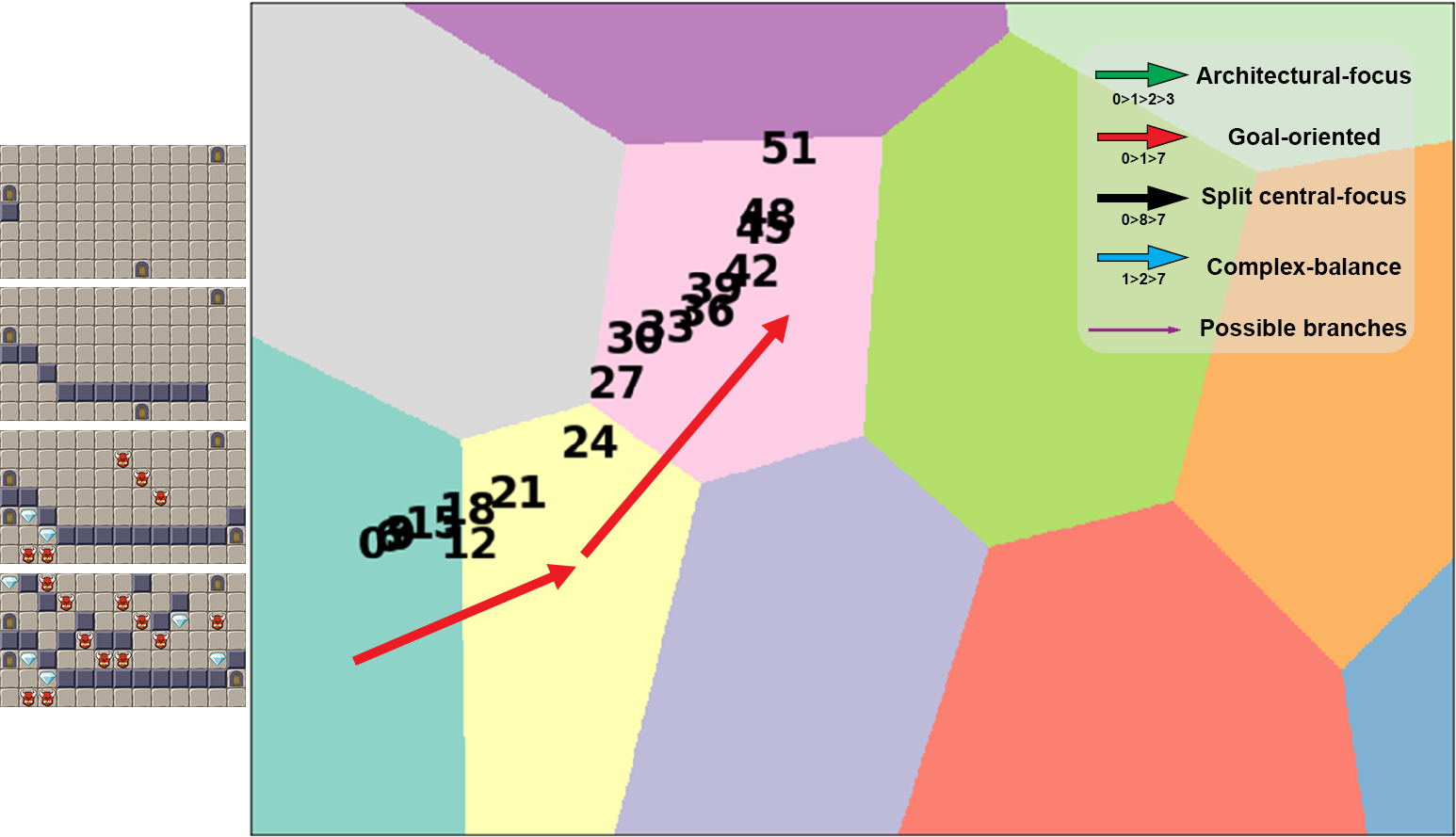}
     }\hfill
    %  \medskip
     \subfloat[\textsc{Split central-focus}\label{fig:scf}]{%
       \includegraphics[width=0.38\textwidth]{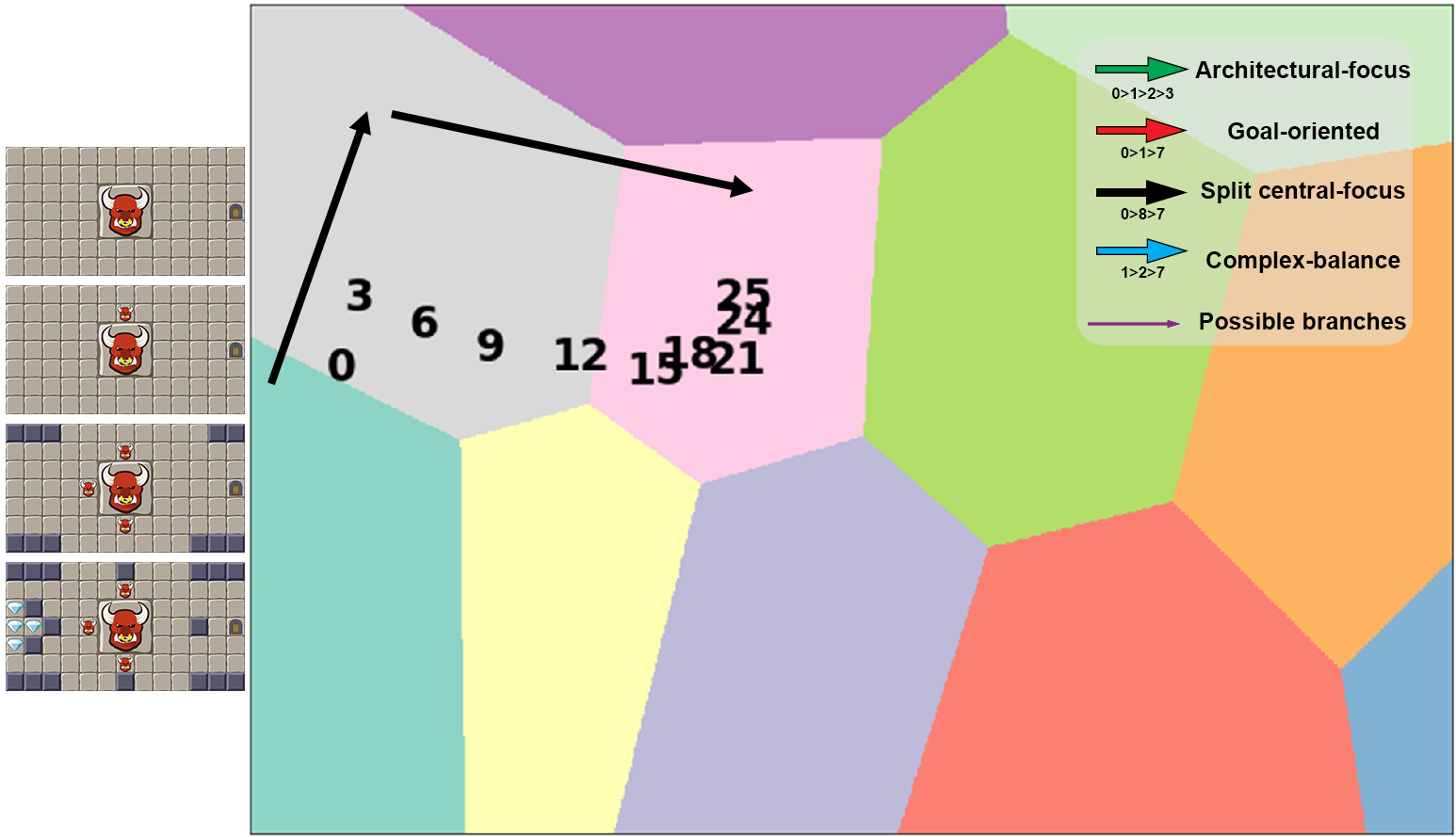}
     }
     \hfill
     \subfloat[\textsc{Complex-balance}\label{fig:cb}]{%
       \includegraphics[width=0.38\textwidth]{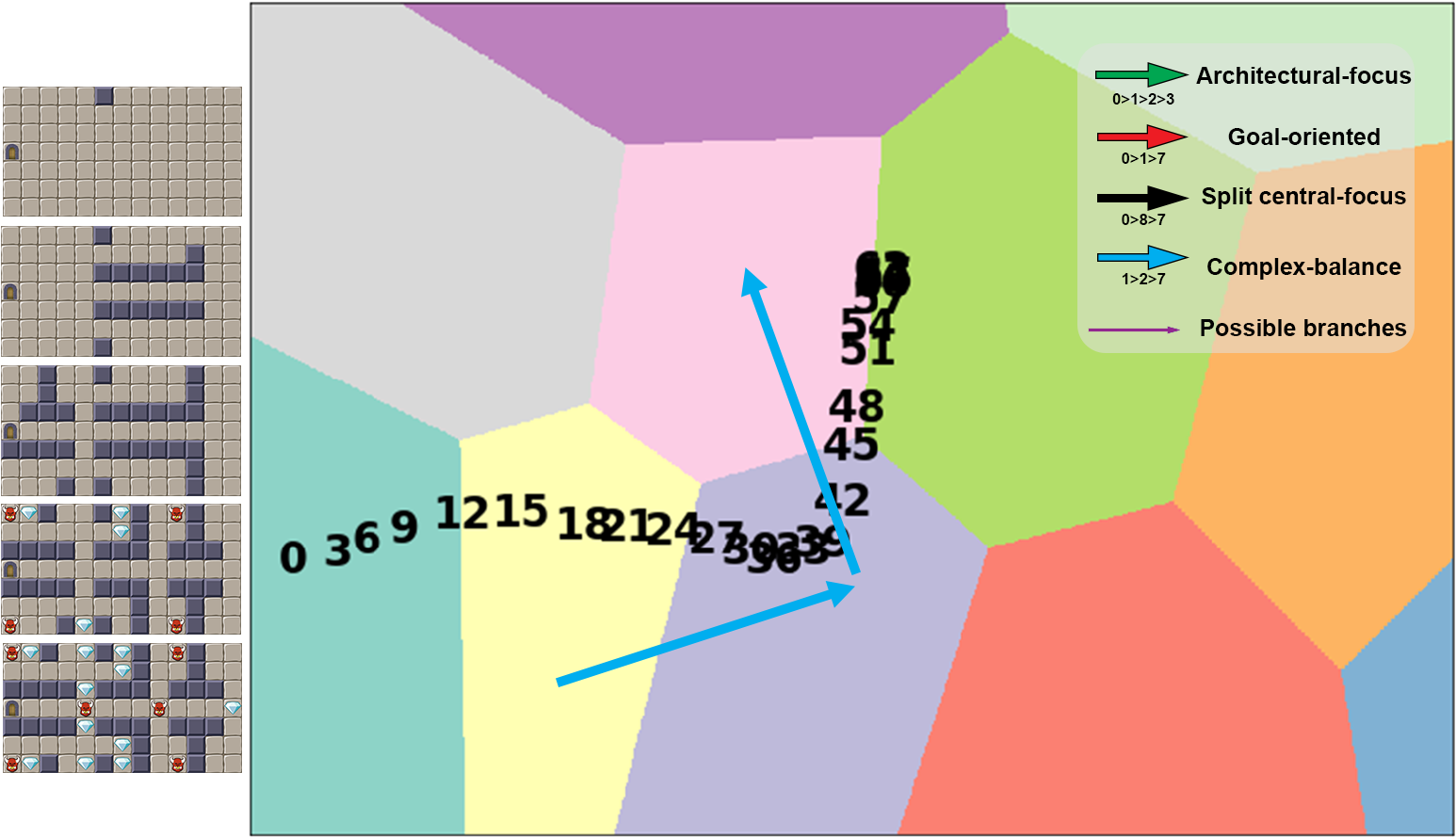}
     }
    
    \caption{Example archetypical paths from different frequent sequences used to create the clusters. To the left of each subfigure, we present each key step in the trajectory i.e. when the design entered a new room style cluster. (a) presents the \textsc{Architectural-focus} where the focus is firstly on creating the structural design of the rooms; the design process jumps back and forth suddenly to room style cluster 5 (one of the possible branches) due to the designer adding a boss, and removing it immediately. (b) presents the \textsc{Goal-oriented} where the design focus on a minimal architecture complexity and mix between adding structural changes and enemies/treasures. (c) shows the \textsc{Split central-focus} where, intentionally, the designer creates a center obstacle with a boss and build around it. Finally, (d) presents the \textsc{Complex-balance}; the designer focuses on building complex, uncommon structures first and then add some goal with enemies and treasures, taking advantage of the spaces.}
    \label{fig:archetypical-examples}
\end{figure}

Once we created, evaluated, and labeled the room style clusters, we were able to cluster and visualize the paths of a typical design session. Figure \ref{fig:paths-designers} presents an example of the design sessions, where we cluster each step of the design. This sequential process revealed that there is an interesting continuity between room style clusters, even capturing when a designer applied one of the procedural suggestions due to bigger steps in the room style clusters. Further, through this process, we could understand the progress of designers in their design process and represent their design style and trajectory in relation to the traversed room style clusters rather than individual editions. 

% In the following sections, we refer to RS-clusters as the room style clusters shown in fig.~\ref{fig:paths-designers} and DS-clusters as 

\subsection{Unique Trajectories}

Using the room style clusters in Figure \ref{fig:all-clusters}, we clustered the design session of all the $180$ designs. This resulted in trajectories where each individual design step was associated to some room style cluster. For instance, the design session in fig.~\ref{fig:paths-designers} resulted the trajectory: $Trajectory=$\{0\textgreater0\textgreater0\textgreater0\textgreater0\textgreater0\textgreater0\textgreater1\textgreater1\textgreater1\textgreater1\textgreater1\textgreater1\textgreater2\textgreater2 \textgreater2\textgreater2\textgreater2\textgreater2\textgreater2\textgreater2\textgreater2\textgreater2\textgreater2\textgreater2\textgreater2\textgreater2\textgreater2\textgreater2\textgreater6\textgreater6\textgreater6\}. We then converted these trajectories to their unique trajectory, which would convert the previous $Trajectory$ into $Unique=$\{0\textgreater1\textgreater2\textgreater6\}, where the first and last element of the sequence are respectively, the starting- and end-points, with all the unique intermediate steps in between.

% in a trajectory where the first $6$ steps were cluster 0, the following 6 were cluster 1, then the  
% and collected the unique trajectories that arose from traversing the various room style clusters. Unique trajectories are identified as the unique steps between room

% These unique trajectories varied in the starting point, length, and end-point, however, when analyzing the trajectories we identified common patterns among them. Using the design session in fig.~\ref{fig:paths-designers}, they had a smiThey had a similar shape as the following, $Unique=$\{0\textgreater8\textgreater7\textgreater6\textgreater10\}, where the first and last element of the sequence are respectively, the starting- and end-points, with all the unique intermediate steps in between.

These unique trajectories varied in the starting point, length, and end-point, however, when analyzing the trajectories we identified common patterns among them. To gather the common patterns from the trajectories, we applied the Generalized Sequential Pattern (GSP) algorithm, which locates frequent subsequences in the analyzed trajectories. For instance, given three trajectories (a) \{8\textgreater1\textgreater2\textgreater6\textgreater9\}, (b) \{8\textgreater1\textgreater2\textgreater6\textgreater10\}, and (c) \{0\textgreater1\textgreater2\textgreater6\}, none of these is a perfect match in its entirety, but GSP can spot that subsequences \{1\textgreater2\textgreater6\}, \{1\textgreater2\}, \{2\textgreater6\}, among others, appear with frequency $= 3$.

%(2) obtain only 1 pattern ($\{5>1>3>11\}$) with frequency=2, if searching from starting points. Finally, using GSP, we find $\{5>1>3>11\}$, $\{1>3>11\}$, $\{1>3\}$, $\{3>11\}$

%We collected these unique trajectories, and 
After doing a preliminary analysis, we identified some steps that we classified as ``border designs'': steps that are borderline between two room style clusters. These \textit{border designs} disrupted the sequence pattern mining with noise in the unique trajectories, specifically when these \textit{border designs} entered a different room style cluster for just a few steps. %we categorize them as when these "unique" noisy steps were brief.
Therefore, we filtered them out by applying a threshold $\theta = 3$, so that all subsequences inside one room style cluster with less than $\theta$ steps are removed from the main sequence. For instance, the sample trajectory \{0\textgreater0\textgreater0\textgreater0\textgreater8\textgreater8\textgreater8\textgreater7\textgreater8\} turns into \{0\textgreater8\} instead of \{0\textgreater8\textgreater7\textgreater8\}. Through this, we were able to reduce the noise and the search space, obtaining meaningful and frequent patterns.

\subsection{Archetypical Paths through Style Space}
\label{sec:archetypical-paths}

%Figure \ref{fig:finalPaths} shows the archetypical paths taken by designers when creating rooms. Represented as arrows to denote direction, 

%From all the collected unique trajectories, we identified 4 main archetypical paths, which are the ones taken most frequently by designers either as their full path or as the initial path. In Figure \ref{fig:finalPaths}, it is shown the archetypical paths, represented as thicker arrows to denote direction, that represent the taken by designers when creating rooms. 

In Figure \ref{fig:finalPaths}, we present the archetypical paths, represented as thicker arrows to denote direction, which show the most frequent paths taken by designers either through their whole design process or as the initial meaningful steps. From all the collected unique trajectories, we have identified 4 main archetypical paths, labelled, \textsc{Architectural-focus}, \textsc{Goal-oriented}, \textsc{Split central-focus}, and \textsc{Complex-balance}. In addition, we have numbered each archetypical path for easier visualization and referencing. In the figure, we also include thinner purple arrows pointing to different room style clusters from several of the room style clusters that are part of the main paths. These are \textit{possible branches} presented in the unique trajectories and added based on their frequency. Through these possible branches, the design of an archetypical session, can vary and extended or deviate the final design. Each archetypical path is defined and explained as follows: 

% \textit{Possible branches} presented in the unique trajectories exhibiting paths deviating from the main path were included as 

\subsubsection{Architectural-focus}The path followed by this archetype focuses first on designing the architecture of the room with walls. Through this, the design focuses on shaping the visual patterns, chambers, and corridors to give a clear space for adding goals and objectives with enemies and treasures. The sequence is denoted with a green arrow in Figure \ref{fig:finalPaths}, and following the sequence \{0\textgreater1\textgreater2\textgreater3\}.

\subsubsection{Goal-oriented}Design processes following this archetypical path, create the rooms in a more standard way, combining simpler symmetric wall structures with distributed placement of enemies and treasures. Thus, rather than focusing extensively on an individual part of the room, the rooms have an initial structure and then they are populated with some specific goal. The sequence is denoted with a red arrow in Figure \ref{fig:finalPaths}, and following the sequence \{0\textgreater1\textgreater7\}.

\begin{figure*}[t!]
    \centering
     \subfloat[Design Goal: Low Leniency\label{fig:lowlen}]{%
       \includegraphics[width=0.3\textwidth]{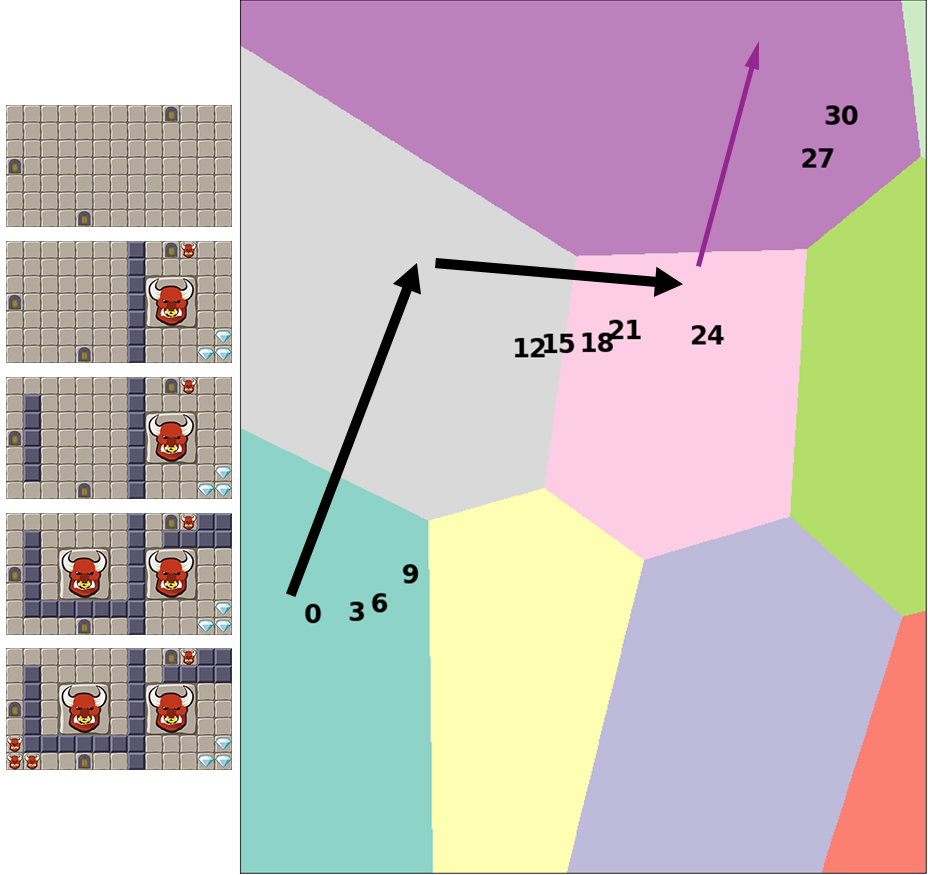}
     }
     \hfill
     \subfloat[Design Goal: High Meso-Patterns\label{fig:highmeso}]{%
       \includegraphics[width=0.3\textwidth]{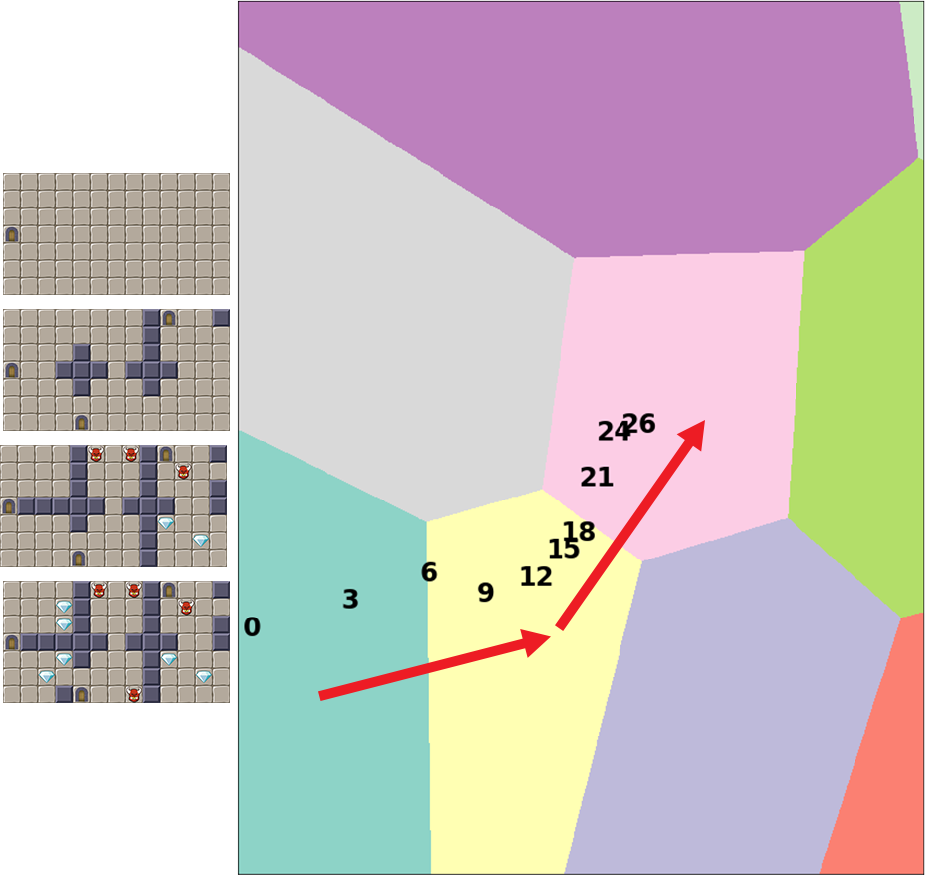}
     }\hfill
    %  \medskip
     \subfloat[Design Goal: Low Linearity\label{fig:lowlin}]{%
       \includegraphics[width=0.3\textwidth]{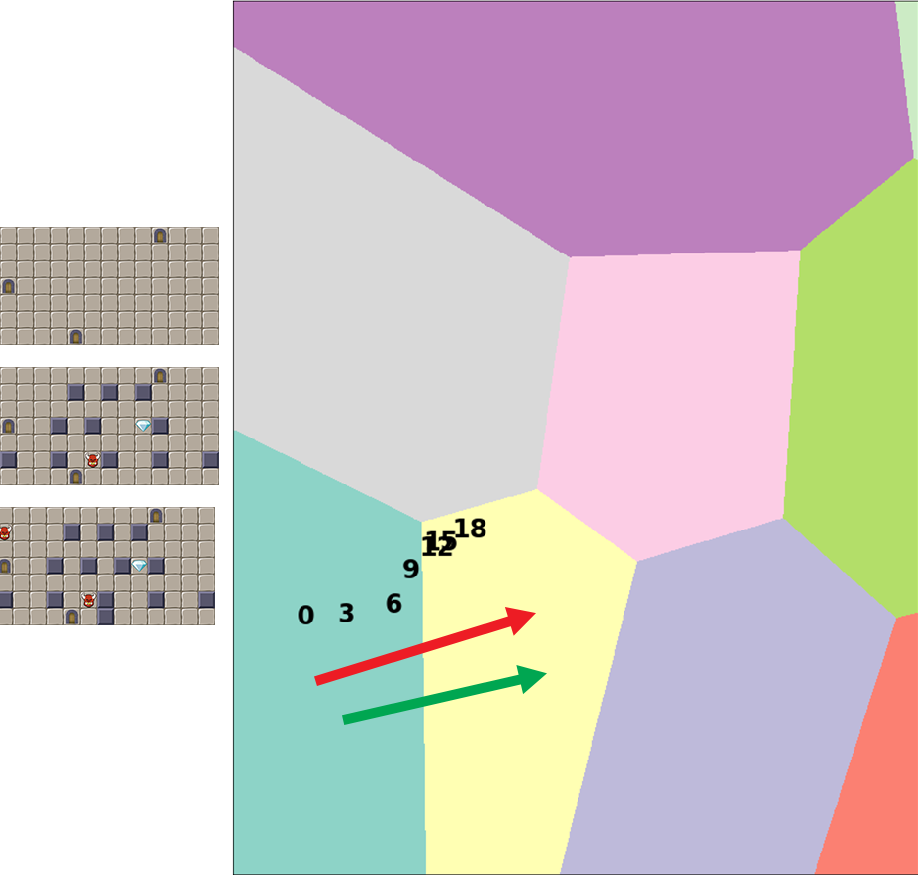}
     }
    
    \caption{Example sequences created with specific design goals orthogonal to the designer personas. (a) and (b) are categorized as \textsc{Split Central-Focus} and \textsc{Goal-Oriented}, respectively. While (c) is uncategorized (with two possible designer personas).}
    \label{fig:prelEval}
\end{figure*}

%Thus, rather than focusing on an individual part of the room until satisfied, the rooms have some initial structures that are populated and continue through an iterative process between these steps.% rooms go through an iterative process of adding have some initial structures that are

\subsubsection{Split central-focus}This archetypical path focuses on designing rooms with obstacles placed in the center of the room in the shape of enemies, treasures, or wall structures that clearly split the room into different areas. The design process is less organized than the other archetypes since it searches to achieve the split goal with any of the available tiles. The sequence is denoted with a black arrow in Figure \ref{fig:finalPaths}, and following the sequence \{0\textgreater8\textgreater7\}.
%, since the middle step is cluster 5 ("Separating and populating chambers"), which relates to rooms which are expected since the cluster 5 ("Separating and populating chambers") relate to rooms that  as specific structural shapes are not necessary. 

\subsubsection{Complex-balance}This archetypical path focuses on building complex symmetric shapes with a clear objective for the player and adapting the spaces with a balance of enemies and treasures. In general, the rooms created following this path are more unique and typically balanced. %   with that adapt well. The process is quite 
The sequence is denoted with a blue arrow in Figure \ref{fig:finalPaths}, and following the sequence \{1\textgreater2\textgreater7\}.

% Furthermore, using these archetypical paths, we can then categorize certain room style clusters as key clusters or being more relevant than others based on their contribution to the paths, their frequency, and their usage. Most of the paths go through or end in cluster 7 (``Balancing and optimizing'') and cluster 1 (``Main architectural patterns''), which relate to rooms that have a more explicit mix between corridors and small chambers, and more clear architecture. The rooms in those clusters are or shaped as end rooms, as in the case of cluster 7, or architecturally shaped to be “optimized” to a specific goal e.g. a dense bordered room. Similarly, most of the sequences start from cluster 0 ("Initial room shapes"), with $134$ out of the $180$ designs, which correlates to the type of designs encountered in that clusters. Thus, it is understandable that most of the archetypical paths pass through these three clusters. 

Furthermore, using these archetypical paths, we can then categorize certain room style clusters as key clusters based on their contribution to the paths, their frequency, and their usage. Most of the paths go through or end in cluster 7 (``Balancing and optimizing'') and cluster 1 (``Main architectural patterns''), which relate to rooms that have a more explicit mix between corridors and small chambers, and more clear architecture. The rooms in those clusters are or shaped as end rooms, as in the case of cluster 7, or architecturally shaped to be “optimized” to a specific goal e.g. a dense bordered room. Similarly, most of the sequences start from cluster 0 ("Initial room shapes"), with $134$ out of the $180$ designs, which correlates to the type of designs encountered in that clusters. Thus, it is understandable that most of the archetypical paths pass through these three clusters. 

Nevertheless, it is the steps in-between that creates a clear differentiation between the archetypical paths, which is the benefit of observing the design process as a whole in the clustered room style space. For instance, in fig.~\ref{fig:finalPaths}, it can be observed that \textsc{Split central-focus} starts in the same room style cluster as three other paths, and tentatively ends in the same room style cluster as three other. However, the designs following \textsc{Split central-focus} are more different from the other trajectories, since it enters a room style cluster that is denser with several tile types in principle, and where designers seem to have a clearer goal.
% With this, we can further understand why \textsc{Split central-focus} is more different to the other trajectories, since it enters a cluster that is "less organized" in principle. 

%Furthermore, we can also observe that certain clusters are key steps for most paths because they are or a frequent starting or ending point. Most of the paths go through or end in cluster 6 ("Balancing and optimizing") and cluster 8 ("Main structural patterns"), which relate to rooms that have a more explicit mix between corridors and small chambers and a more clear structure, thus, it is understandable since the rooms in those clusters are or shaped as end rooms, as in the case of cluster 6, or structurally shaped to be “optimized” to a specific goal (E.g. dense bordered room, maze-like, more challenging, etc.). However, the distribution of endpoints is quite even, and meanwhile, cluster 6 and cluster 11 ("Chamber separation, Forced enemy encounter") are the most frequent ending points with $36$ and $25$ out of $180$ design processes, the rest of clusters are quite close.

%Similarly, most of the sequences start from cluster 0 ("Initial room shapes"), with $134$ out of the $180$ designs, which correlates to the type of designs encountered in those clusters.

Moreover, in figure~\ref{fig:archetypical-examples}, we present examples of each of the designer personas by visualizing the sequence of steps done in representative design sessions, showing what these paths would look like in practice. Each visualization of a designer persona has the key design steps to the left, where each image is in a sequence: the first is the first edition of the designer, the last is the final edition, and the in-between represent entering a new room style cluster. 

Figure~\ref{fig:af} shows the \textsc{Architectural-focus}, where the designer first created the border of the room with a clear chamber division. As the designer adds and subsequently removes the boss, the design jumps to cluster 10, which is one of the possible branches, adding a high challenge. 

Figure~\ref{fig:go} shows the \textsc{Goal-Oriented}, where the designer sketched the main shape of the room followed by alternating between enemies, treasures, and walls to adjust the room towards the goal for the player. In this example, the designer ends the design close to cluster 9, with a disorganized placement of tiles and a less aesthetical room, but forming small choke areas balancing the placement of enemies and treasures.

Figure~\ref{fig:scf} shows the \textsc{Split Central-focus}, where the designer directly started by adding a boss in the center of the room and using this as a reference point, shaped the rest of the room. Figure~\ref{fig:cb} shows the \textsc{Complex-balance}, where the designer focused on creating an uncommon structure and followed by adding enemies and treasures symmetrically, with clear individual areas for the player to approach.

% It is not surprising to focus on the center as it 

Finally, further analyzing figure~\ref{fig:archetypical-examples}, it can also be observed an interesting dual tendency of the designers in the archetypical paths. This dual tendency is to either focus on the aesthetic configuration of the room based on what is perceived in the editor exemplified the personas: \textsc{Architectural-focus} and \textsc{Split central-focus}, and to focus on the player experience exemplified the personas: \textsc{Goal-oriented} and \textsc{Complex-balance}. Nevertheless, both are not mutually exclusive, instead this illustrates adequately the dualistic role the designer has when using the tool and designing rooms. That of creating an aesthetically pleasing object as it is seen in the editor, and that of creating an experience.% However, this is not mutually exclusive. Instead, it shows 

% \begin{figure*}[t!]
% \centerline{\includegraphics[width=16cm]{figures/new_evalaution-3-seq_other.png}}
% \caption{Preliminary evaluation of the systems} 
% \label{fig:prelEval}
% \end{figure*}

\subsection{Preliminary Evaluation}

We recorded three design sessions with specific design goals orthogonal to the archetypical trajectories to preliminary evaluate our approach. These are shown in figure~\ref{fig:prelEval} with the design goals: figure~\ref{fig:lowlen}) Low Leniency (i.e., challenging room), figure~\ref{fig:highmeso}) high meso-pattern level (i.e., large quantity of guard, treasure, enemy, and ambush rooms), and figure~\ref{fig:lowlin}) low linearity (i.e., several paths between doors). Both figure~\ref{fig:lowlen} and figure~\ref{fig:highmeso} fitted two of the archetypical trajectories, \textsc{Split Central-Focus} and \textsc{Goal-Oriented}. Figure~\ref{fig:lowlin} shows a design achieving its goal but identified as an early design, aligned with two possible designer personas. This last example shows the situations we expect to normally encounter during design sessions, namely, uncertainty and adaptation. If the designer continued elaborating the level's architecture, they would align with the \textsc{Architectural Focus} else they would probably align with the \textsc{Goal-Oriented}.

While these results requires a larger study that analyses and verifies the different possibilities and properties of using Designer Personas, especially, in-the-wild; this evaluation helps to understand and visualize how unseen design sessions map and align to possible archetypical paths regardless of the designer's design goal.

\section{Discussion}

Our work draws on many of Liapis et al.'s~\cite{Liapis2013-designerModel,Liapis2014-designerModelImpl} ideas, concepts, and goals but differs on the tool and type of game being created and the methods to create designer models. These differences strengthen the importance and usefulness of designer modeling and highlight this designer-centric perspective's holistic and generic properties.

The archetypical paths presented in sec.~\ref{sec:archetypical-paths} relate to EDD and the options that exist within it. However, while important for EDD's development, this is used to demonstrate what can be done and as an exemplification of the method to model designers. Our method focuses on designer modeling through clustering the design space and the room style based on level design step sequences, resulting in archetypical paths. We leverage a collective dataset to build a collective cluster model of a set of designers and utilize this to mitigate, to some extent, problems regarding lack of individual designer data and continuous model adaptation~\cite{Alvarez2020-DesignerPreference}. Our method allows us to better understand, cluster, categorize and isolate designer behavior. A virtual model of the designer's style could allow to better drive the search process for procedurally generating content that is valuable for designers and aligned with the designers' intentions, observed as well by Liapis et al.~\cite{Liapis2013-designerModel}.

%that is valuable for and aligned with the designer's intentions as discussed by Liapis et al..

%This could be very valuable for mixed-initiative approaches, where a clear virtual model of the designer's style allows us to better drive the search process for procedurally generating content that is valuable for and aligned with the designer's intentions.

We expect that work on similar domains and tools working with level design could discuss these archetypical paths in relation to what designers create. They can build on and adapt the proposed method to model designers to their respective domains, identifying valuable and applicable designer personas. Then, these could be used to understand what designers create and to design adaptive systems and better Human-AI collaborations and interactions. The resulting Designer Personas have the potential to be used in many different scenarios. For instance, as objectives for a search-based approach to enable a more style-sensitive system, to evaluate the fitness of evolutionary generated content, or to train PCG agents via Reinforcement Learning~\cite{khalifa2020-pcgrl}.

\section{Conclusions}

% \begin{itemize}
%     \item Discussion on what does this archetypical design trajectories mean?
%     \item how to use them? next steps into integrating this into a system. To use this in a search-based approach as objectives for the generation to move towards the directions where (according to our archetypical design trajectories) the designers will move towards in their design process. Perhaps I could also bring the discussion from the workshop-paper for HC-AI.
%     \item discussion on creativity? is the output or the process where the actual creativity is outputted? Compare using end-design clustering to using sequences to cluster.
%     \item Discussion on how PCGRL relates to this type of work? --> Perhaps this is something for the background instead.
% \end{itemize}{}

% This paper presents a step towards designer modelling in a MI-CC environment by providing an implementation of designer personas as archetypical trajectories through style space, as a means to characterize several representative and frequent design styles together. 

%This paper presents a novel approach and meaningful steps towards designer modeling in an MI-CC environment. By providing an implementation of designer personas as archetypical trajectories through style space, we show that 

This paper presents a novel method and meaningful steps towards designer modeling through an experiment on archetypical design trajectories analysis in an MI-CC environment. Through this, we characterize several representative design styles as designer personas. We have first run and compared several clustering setups to find the best partitioning of the room style using the design step sequence of the collected $180$ unique rooms, ending in $8196$ data points, and resulting in a set of twelve cohesive and representative room style clusters. We have then mapped these $180$ design sequences in terms of these clusters, applying frequent sequence mining to find four frequent and unique designer styles, with related common sub-styles. As a result, we have presented a roadmap of design styles over a map of data-driven room style clusters.

Recognizing the designers' current style and the path taken so far, which would indicate a possible designer persona, would open the possibility for recognizing their intentions, preferences, and goals. This traced roadmap of designer personas could let a content generator anticipate a designer's next moves without heavy computational cost, just by identifying their current location on the map and offering content suggestions that lie in the most promising clusters to be visited next. Conversely, it would also identify designers who do not follow a certain path, i.e., deviating from the pattern, trying to understand their objective through their design style. Finally, we aim at implementing our approach in a functional system within EDD to assess and validate the benefits and usability of these adaptable systems with human designers.

\bibliographystyle{IEEEtran}
\bibliography{references.bib}

\begin{thebibliography}{10}
\providecommand{\url}[1]{#1}
\csname url@samestyle\endcsname
\providecommand{\newblock}{\relax}
\providecommand{\bibinfo}[2]{#2}
\providecommand{\BIBentrySTDinterwordspacing}{\spaceskip=0pt\relax}
\providecommand{\BIBentryALTinterwordstretchfactor}{4}
\providecommand{\BIBentryALTinterwordspacing}{\spaceskip=\fontdimen2\font plus
\BIBentryALTinterwordstretchfactor\fontdimen3\font minus
  \fontdimen4\font\relax}
\providecommand{\BIBforeignlanguage}[2]{{%
\expandafter\ifx\csname l@#1\endcsname\relax
\typeout{** WARNING: IEEEtran.bst: No hyphenation pattern has been}%
\typeout{** loaded for the language `#1'. Using the pattern for}%
\typeout{** the default language instead.}%
\else
\language=\csname l@#1\endcsname
\fi
#2}}
\providecommand{\BIBdecl}{\relax}
\BIBdecl

\bibitem{yannakakis2014micc}
G.~N. Yannakakis, A.~Liapis, and C.~Alexopoulos, ``Mixed-initiative
  co-creativity,'' in \emph{Proceedings of the 9th Conference on the
  Foundations of Digital Games}, 2014.

\bibitem{liapis2016mixed}
A.~Liapis, G.~Smith, and N.~Shaker, ``Mixed-initiative content creation,'' in
  \emph{Procedural content generation in games}.\hskip 1em plus 0.5em minus
  0.4em\relax Springer, 2016, pp. 195--214.

\bibitem{LUBART2005-computerPartners}
\BIBentryALTinterwordspacing
T.~Lubart, ``How can computers be partners in the creative process:
  Classification and commentary on the special issue,'' \emph{International
  Journal of Human-Computer Studies}, vol.~63, no.~4, pp. 365 -- 369, 2005,
  computer support for creativity. [Online]. Available:
  \url{http://www.sciencedirect.com/science/article/pii/S1071581905000418}
\BIBentrySTDinterwordspacing

\bibitem{Holmgard2019-proceduralPersonas}
C.~{Holmgård}, M.~C. {Green}, A.~{Liapis}, and J.~{Togelius}, ``Automated
  playtesting with procedural personas through mcts with evolved heuristics,''
  \emph{IEEE Transactions on Games}, vol.~11, no.~4, pp. 352--362, Dec 2019.

\bibitem{tloz}
{Nintendo R\&D4}, ``{The Legend of Zelda},'' 1986.

\bibitem{Togelius2011}
J.~{Togelius}, G.~N. {Yannakakis}, K.~O. {Stanley}, and C.~{Browne},
  ``Search-based procedural content generation: A taxonomy and survey,''
  \emph{IEEE Transactions on Computational Intelligence and AI in Games},
  vol.~3, no.~3, pp. 172--186, Sep. 2011.

\bibitem{khalifa2020-pcgrl}
A.~Khalifa, P.~Bontrager, S.~Earle, and J.~Togelius, ``Pcgrl: Procedural
  content generation via reinforcement learning,'' 2020.

\bibitem{alvarez2019empowering}
A.~Alvarez, S.~Dahlskog, J.~Font, and J.~Togelius, ``Empowering quality
  diversity in dungeon design with interactive constrained map-elites,'' in
  \emph{2019 IEEE Conference on Games (CoG)}, 2019.

\bibitem{Baldwin2017}
A.~Baldwin, S.~Dahlskog, J.~M. Font, and J.~Holmberg, ``Towards pattern-based
  mixed-initiative dungeon generation,'' in \emph{Proceedings of the 12th
  International Conference on the Foundations of Digital Games}, ser. FDG
  '17.\hskip 1em plus 0.5em minus 0.4em\relax New York, NY, USA: ACM, 2017.

\bibitem{thawonmas2019artificial}
R.~Thawonmas, J.~Togelius, and G.~N. Yannakakis, ``Artificial general
  intelligence in games: Where play meets design and user experience,''
  \emph{NII Shonan Meeting No. 130}, 2019.

\bibitem{melhart2020feel}
D.~Melhart, G.~N. Yannakakis, and A.~Liapis, ``I feel i feel you: A theory of
  mind experiment in games,'' \emph{KI-K{\"u}nstliche Intelligenz}, pp. 1--11,
  2020.

\bibitem{Melhart2019-ModellingMotivation}
D.~{Melhart}, A.~{Azadvar}, A.~{Canossa}, A.~{Liapis}, and G.~N. {Yannakakis},
  ``Your gameplay says it all: Modelling motivation in tom clancy’s the
  division,'' in \emph{2019 IEEE COG}, Aug 2019, pp. 1--8.

\bibitem{canossa2015towards}
A.~Canossa and G.~Smith, ``Towards a procedural evaluation technique: Metrics
  for level design,'' in \emph{The 10th International Conference on the
  Foundations of Digital Games}.\hskip 1em plus 0.5em minus 0.4em\relax sn,
  2015, p.~8.

\bibitem{Alvoz2019-PersonalityDriven}
A.~Alvarez and M.~Vozaru, ``Perceived behaviors of personality-driven agents,''
  \emph{Violence | Perception | Video Games: New Directions in Game Research},
  pp. 171--184, 2019.

\bibitem{Drachen2009-playerModellingTombRaider}
A.~{Drachen}, A.~{Canossa}, and G.~N. {Yannakakis}, ``Player modeling using
  self-organization in tomb raider: Underworld,'' in \emph{2009 IEEE Symposium
  on Computational Intelligence and Games}, Sep. 2009, pp. 1--8.

\bibitem{summerville2018procedural}
A.~Summerville, S.~Snodgrass, M.~Guzdial, C.~Holmg{\aa}rd, A.~K. Hoover,
  A.~Isaksen, A.~Nealen, and J.~Togelius, ``Procedural content generation via
  machine learning (pcgml),'' \emph{IEEE Transactions on Games}, vol.~10,
  no.~3, pp. 257--270, 2018.

\bibitem{Duque2021-BayesianbasedPlayerModel}
M.~G. Duque, R.~B. Palm, and S.~Risi, ``Fast game content adaptation through
  bayesian-based player modelling,'' in \emph{IEEE Conference on Games (CoG)},
  2021.

\bibitem{togelius2007-AutomaticPersonalisedRaceGames}
J.~Togelius, R.~De~Nardi, and S.~M. Lucas, ``Towards automatic personalised
  content creation for racing games,'' in \emph{2007 IEEE Symposium on
  Computational Intelligence and Games}, 2007, pp. 252--259.

\bibitem{Yannakakis2011-experiencedrivenPCG}
G.~N. {Yannakakis} and J.~{Togelius}, ``Experience-driven procedural content
  generation,'' \emph{IEEE Transactions on Affective Computing}, vol.~2, no.~3,
  pp. 147--161, 2011.

\bibitem{Summerville2016-LearningPlayerTailoredPlatformer}
\BIBentryALTinterwordspacing
A.~Summerville, M.~Guzdial, M.~Mateas, and M.~Riedl, ``Learning player tailored
  content from observation: Platformer level generation from video traces using
  lstms,'' \emph{Proceedings of the AAAI Conference on Artificial Intelligence
  and Interactive Digital Entertainment}, vol.~12, no.~2, pp. 107--113, 2016.
  [Online]. Available:
  \url{https://ojs.aaai.org/index.php/AIIDE/article/view/12895}
\BIBentrySTDinterwordspacing

\bibitem{Panagiotis2021-susketch}
P.~Migkotzidis and A.~Liapis, ``Susketch: Surrogate models of gameplay as a
  design assistant,'' \emph{IEEE Transactions on Games}, pp. 1--1, 2021.

\bibitem{charity2020baba}
M.~Charity, A.~Khalifa, and J.~Togelius, ``Baba is y'all: Collaborative
  mixed-initiative level design,'' \emph{arXiv: 2003.14294}, 2020.

\bibitem{machado2019pitako}
T.~Machado, D.~Gopstein, A.~Nealen, and J.~Togelius, ``Pitako-recommending game
  design elements in cicero,'' in \emph{2019 IEEE Conference on Games
  (CoG)}.\hskip 1em plus 0.5em minus 0.4em\relax IEEE, 2019.

\bibitem{shaker2013ropossum}
N.~Shaker, M.~Shaker, and J.~Togelius, ``Ropossum: An authoring tool for
  designing, optimizing and solving cut the rope levels.'' in \emph{AIIDE},
  2013.

\bibitem{smith_tanagra:_2011}
G.~Smith, J.~Whitehead, and M.~Mateas, ``Tanagra: {Reactive} {Planning} and
  {Constraint} {Solving} for {Mixed}-{Initiative} {Level} {Design},''
  \emph{IEEE Transactions on Computational Intelligence and AI in Games},
  vol.~3, no.~3, Sep. 2011.

\bibitem{liapis_generating_2013}
A.~Liapis, G.~N. Yannakakis, and J.~Togelius, ``Generating {Map} {Sketches} for
  {Strategy} {Games},'' in \emph{Proceedings of {Applications} of
  {Evolutionary} {Computation}}, vol. 7835, LNCS.\hskip 1em plus 0.5em minus
  0.4em\relax Springer, 2013.

\bibitem{Liapis2013-designerModel}
A.~Liapis, G.~Yannakakis, and J.~Togelius, ``\BIBforeignlanguage{English
  (US)}{Designer modeling for personalized game content creation tools},'' in
  \emph{\BIBforeignlanguage{English (US)}{Artificial Intelligence and Game
  Aesthetics - Papers from the 2013 AIIDE Workshop, Technical Report}}, vol.
  WS-13-19.\hskip 1em plus 0.5em minus 0.4em\relax AI Access Foundation, 2013,
  pp. 11--16.

\bibitem{Liapis2014-designerModelImpl}
A.~{Liapis}, G.~N. {Yannakakis}, and J.~{Togelius}, ``Designer modeling for
  sentient sketchbook,'' in \emph{2014 IEEE Conference on Computational
  Intelligence and Games}, Aug 2014, pp. 1--8.

\bibitem{adadi2018peeking}
A.~Adadi and M.~Berrada, ``Peeking inside the black-box: A survey on
  explainable artificial intelligence (xai),'' \emph{IEEE Access}, vol.~6, pp.
  52\,138--52\,160, 2018.

\bibitem{Doshi-Velez2018}
F.~Doshi-Velez and B.~Kim, ``{Considerations for Evaluation and Generalization
  in Interpretable Machine Learning},'' in \emph{Explainable and Interpretable
  Models in Computer Vision and Machine Learning}, H.~J. Escalante,
  S.~Escalera, I.~Guyon, X.~Bar{\'{o}}, Y.~G{\"{u}}{\c{c}}l{\"{u}}t{\"{u}}rk,
  U.~G{\"{u}}{\c{c}}l{\"{u}}, and M.~van Gerven, Eds.\hskip 1em plus 0.5em
  minus 0.4em\relax Cham: Springer International Publishing, 2018, pp. 3--17.

\bibitem{Nourani2019-meaningfulExplanations}
S.~M. Mahsan~Nourani, Samia~Kabir and E.~D. Ragan, ``The effects of meaningful
  and meaningless explanations on trust and perceived system accuracy in
  intelligent systems,'' in \emph{Proceedings of the AAAI Conference on Human
  Computation and Crowdsourcing}, 2019.

\bibitem{Zhu2018-XAIDesignersMICC}
J.~{Zhu}, A.~{Liapis}, S.~{Risi}, R.~{Bidarra}, and G.~M. {Youngblood},
  ``Explainable ai for designers: A human-centered perspective on
  mixed-initiative co-creation,'' in \emph{2018 IEEE Conference on
  Computational Intelligence and Games (CIG)}, Aug 2018, pp. 1--8.

\bibitem{guzdial-lvldsg-aiide-2018}
M.~Guzdial, N.~Liao, and M.~Riedl, ``Co-creative level design via machine
  learning,'' in \emph{Joint Proceedings of the {AIIDE} 2018 Workshops
  co-located with 14th {AAAI} Conference on Artificial Intelligence and
  Interactive Digital Entertainment {(AIIDE} 2018), Edmonton, Canada, November
  13-14, 2018.}, 2018.

\bibitem{guzdial2019friend}
M.~Guzdial, N.~Liao, J.~Chen, S.-Y. Chen, S.~Shah, V.~Shah, J.~Reno, G.~Smith,
  and M.~O. Riedl, ``Friend, collaborator, student, manager: How design of an
  ai-driven game level editor affects creators,'' in \emph{Proceedings of the
  2019 CHI Conference on Human Factors in Computing Systems}, 2019, pp. 1--13.

\bibitem{Alvarez2018}
A.~Alvarez, S.~Dahlskog, J.~Font, J.~Holmberg, C.~Nolasco, and A.~\"{O}sterman,
  ``Fostering creativity in the mixed-initiative evolutionary dungeon
  designer,'' in \emph{Proceedings of the 13th International Conference on the
  Foundations of Digital Games}, ser. FDG '18, 2018.

\bibitem{Khalifa2018}
A.~Khalifa, S.~Lee, A.~Nealen, and J.~Togelius, ``Talakat: Bullet hell
  generation through constrained map-elites,'' in \emph{Proceedings of The
  Genetic and Evolutionary Computation Conference}.\hskip 1em plus 0.5em minus
  0.4em\relax ACM, 2018.

\bibitem{Alvarez2020-DesignerPreference}
A.~Alvarez and J.~Font, ``Learning the designer's preferences to drive
  evolution,'' in \emph{Proceedings of the 23rd European Conference on the
  Applications of Evolutionary and bio-inspired Computation}, ser.
  EvoApplications '20, 2020.

\bibitem{mcmillen_binding_2011}
E.~McMillen and F.~Himsl, ``The {Binding} of {Isaac},'' 2011.

\bibitem{Alvarez2020-ICMAPE}
A.~Alvarez, S.~Dahlskog, J.~Font, and J.~Togelius, ``Interactive constrained
  map-elites: Analysis and evaluation of the expressiveness of the feature
  dimensions,'' \emph{arXiv: 2003.03377}, 2020.

\bibitem{scikit-learn}
F.~Pedregosa, G.~Varoquaux, A.~Gramfort, V.~Michel, B.~Thirion, O.~Grisel,
  M.~Blondel, P.~Prettenhofer, R.~Weiss, V.~Dubourg, J.~Vanderplas, A.~Passos,
  D.~Cournapeau, M.~Brucher, M.~Perrot, and E.~Duchesnay, ``Scikit-learn:
  Machine learning in {P}ython,'' \emph{Journal of Machine Learning Research},
  vol.~12, pp. 2825--2830, 2011.

\end{thebibliography}

\ifCLASSOPTIONcaptionsoff
  \newpage
\fi

\end{document}